\documentclass[12pt, draftclsnofoot, onecolumn]{IEEEtran}
\usepackage{graphicx, times, amsmath, amsfonts, comment,amsthm}
\usepackage{amssymb, epstopdf, soul, color}
\usepackage[noend]{algorithmic}
\usepackage{algorithm, array}

\newcommand{\beq}{\begin{equation}}
\newcommand{\eeq}{\end{equation}}
\newcommand{\tbf}{\textbf}
\newcommand{\tit}{\textit}

\newcommand{\argmin}{\operatornamewithlimits{argmin}}

\newcommand{\ud}{\mathrm{d}}
\newcommand{\sgn}{\mathrm{sgn}}

\theoremstyle{plain}
\newtheorem{lemmacounter}{Lemma}
\newtheorem{lemma}[lemmacounter]{Lemma}
\theoremstyle{plain}
\newtheorem{theoremcounter}{Theorem}
\newtheorem{theorem}[theoremcounter]{Theorem}
\theoremstyle{plain}
\newtheorem{propcounter}{Proposition}
\newtheorem{proposition}[propcounter]{Proposition}
\theoremstyle{plain}
\newtheorem{corocounter}{Corollary}
\newtheorem{corollary}[corocounter]{Corollary}
\theoremstyle{plain}

\newcommand {\Ebb}{\mathbb{E}}
\newcommand {\Kbb}{\mathbb{K}}
\newcommand {\Sbb}{\mathbb{S}}
\newcommand {\Vbb}{\mathbb{V}}

\newcommand {\Rbb}{\mathbb{R}}

\newcommand {\Bcal}{\mathcal{B}}

\newcommand {\Kcal}{\mathcal{K}}
\newcommand {\Lcal}{\mathcal{L}}
\newcommand {\Mcal}{\mathcal{M}}

\begin{document}
\title{SINR Outage Evaluation in Cellular Networks: Saddle Point Approximation (SPA) Using Normal Inverse Gaussian (NIG) Distribution 
}

\author{
\IEEEauthorblockN{Sudarshan Guruacharya, Hina Tabassum, and Ekram Hossain\thanks{The authors are with the Department of Electrical and Computer Engineering at the University of Manitoba, Canada (emails: \{Sudarshan.Gurucharya, Hina.Tabassum, Ekram.Hossain\}@umanitoba.ca).}}
}
\maketitle


\begin{abstract}
Signal-to-noise-plus-interference ratio (SINR) outage probability is among one of the key performance metrics of a wireless cellular network.  
In this paper, we propose  a semi-analytical method based on saddle point  approximation (SPA) technique to calculate the SINR outage of a wireless system whose SINR can be modeled in the form $\frac{\sum_{i=1}^M X_i}{1+\sum_{i=1}^N Y_i}$ where $X_i$ denotes the useful signal power, $Y_i$ denotes the power of the interference signal. Both $M$ and $N$ can also be random variables.  The proposed approach is based on the saddle point approximation to cumulative distribution function (CDF) as given by \tit{Wood-Booth-Butler formula}. The approach is applicable whenever the cumulant generating function (CGF) of the received signal and interference exists, and it allows us to tackle distributions with large skewness and kurtosis with higher accuracy.  In this regard, we exploit a  four parameter \tit{normal-inverse Gaussian} (NIG) distribution as a base distribution.  Given that the skewness and kurtosis satisfy a specific condition, NIG-based SPA works reliably.  When this condition is violated, we recommend SPA based on normal or symmetric NIG distribution, both special cases of NIG distribution, at the expense of reduced accuracy. For the purpose of demonstration, we apply SPA for the SINR outage evaluation of a typical user experiencing a downlink coordinated multi-point transmission (COMP) from the base stations (BSs) that are modeled by homogeneous Poisson point process.   
Numerical results are presented to illustrate the accuracy of the proposed set of approximations.
\end{abstract}




\section{Introduction}
Signal-to-noise-plus-interference ratio (SINR) outage probability (i.e., the probability of the SINR falling below a predefined threshold)  is one of the primary performance metrics for the analysis of a wireless communication system. The simplicity of its definition as well as its connection with other performance parameters, such as bit/symbol error rate and ergodic capacity, make it of significant interest to researchers.

\subsection{Background Work}

Till date, there have been numerous researches that have focused on analyzing the exact SINR outage probability of wireless systems in diverse network settings and under varying modeling assumptions such as random multi-path channel fading \cite[references therein]{Annamalai2001}, aggregate interference, and random distance geometry of transmitters/receivers. 
A popular and powerful approach  is to utilize the moment generating function (MGF) of the received signal and interference random variables. 
Provided the MGF of the received signal power and aggregate interference, \tit{Gil-Pelaez's inversion formula} for cumulative distribution function (CDF) can be applied to find the exact SINR outage at a receiver~\cite{Gil-Pelaez1951}. This involves calculating an integral involving  the MGFs of received signal power and aggregate interference. The integral is typically too complicated to be solved in closed-form. As such, either the integral is expressed in terms of special functions (whose stability may  be unknown)  or  is evaluated using numerical integration techniques. 

It is therefore worthwhile to examine how outage can be approximated to avoid explicit numerical integrations as much as possible. 
One such technique, which was investigated in the late 20th century, is based on saddle point approximation (SPA) method \cite{LugannaniRice1980, Daniels1987, WoodBoothButler1993}. Readers are referred to \cite{Wong2001,Butler2007} for general introduction to saddle point method and its use in statistics. It involves evaluating the cumulant generating function (CGF) at a single point, called the saddle point, where most of the value of integral is concentrated. Saddle point method has previously been utilized  to evaluate detection probability \cite{Helstrom1978, Farshad2011}, error probability \cite{Yue1979}, coding and quantization \cite{Martinez2011,Nevat2013,Xue2014}, and estimation \cite{Beierholm2008}. Recently, in \cite{Guruacharya2016}, we exploited the SPA based on \tit{Lugannani-Rice formula} to compute the SINR outage probability in a scenario where both the desired channel and the interfering channels are modeled using multi-path fading distributions (e.g., Nakagami-$m$, Nakagami-$q$, and Rician). The number of interferers and the distance geometry between the receivers and the transmitters were  deterministic. In \cite{Guruacharya2016a}, we derived a few approximations for coverage probability, but they cannot be extended to other scenarios.

\subsection{Contributions}

The outage analysis of the  cellular networks with random multi-path fading channels and random locations/number of the interferers is  more complicated due to different sources of uncertainty.  Stochastic geometry is often used to analyze such systems (see \cite{Haenggi2012} for a general introduction). In this case, the interference distributions are often intractable and tend to have high skewness and kurtosis (heavy tailed). Consequently, well-known approximations like Lugannani-Rice formula may not provide accurate performance characterization of the system. The Lugannani-Rice formula is found to be accurate when the distribution to be approximated is nearly Gaussian. For highly skewed or heavy tailed distributions, Lugannani-Rice formula  yields inaccurate outage results (e.g., results being negative or greater than unity)~\cite{BoothWood1994, Annaert2007}. Such cases highlight the importance of including the higher order moments in the SPA approximation. This is exactly the case when stochastic geometry is used to model interference. For example, the interference experienced by a typical user is known to follow an alpha-stable distribution when the interfering base stations are modeled as homogeneous Poisson point process \cite{Ilow1998}. 

In this context, the contributions of this paper can be summarized as follows:
\begin{itemize}
\item This paper proposes an  approximation for SINR outage probability based on saddle point methods.  In particular, we propose a general version of the saddle point method introduced by Wood, Booth and Butler~\cite{WoodBoothButler1993}, from which the \tit{Lugannani-Rice formula} can be derived as a special case when normal distribution is considered as the base distribution. The technique utilizes the CGF of the signal and interference variables and allows us to tackle distributions with heavy skewness and tails.

\item We use a  four parameter \tit{normal-inverse Gaussian} (NIG) distribution as a base distribution. NIG offers a higher flexibility in adjusting the shape of the distribution and the decay rates of the tail. To the best of our knowledge, NIG has not been used in the context of SPA. The use of NIG distribution as the base distribution for SPA is in itself of considerable novelty and usefulness, not just for the study of communication systems, but also for statistics in general, other branches of engineering, science, finance, and economics.

\item The proposed approach is general to calculate the SINR outage of any  wireless system whose SINR can be modeled as $\frac{\sum_{i=1}^M X_i}{1+\sum_{i=1}^N Y_i}$ where $X_i$ denotes the $i$-th component of received signal power and $Y_i$ denotes the $i$-th component of interference power. The $M$ and $N$ can also be random variables. For demonstration purposes, we  consider a stochastic geometry-based cellular network model, where base stations (BSs) are distributed according to a homogeneous Poisson Point Process (PPP) and  a typical user experiences a downlink coordinated multi-point (COMP) transmission.  

\item Numerical results show the higher accuracy of the  NIG-based SPA over normal-based SPA (as given by Lugannani-Rice formula). Given that the skewness and kurtosis satisfy specific conditions (referred as {\em sufficient conditions} in this paper), NIG-based SPA works reliably. However, when the specific conditions are not fulfilled, we recommend SPA based on normal distribution or on symmetric NIG distribution, at the expense of reduced accuracy.
\end{itemize}

\subsection{Paper Organization and Notations}

The rest of the paper is structured as follows: Section~II discusses the SINR model along with the basic definitions we will be using throughout the paper
and the typical application of Gil-Pelaez formula for outage computation. Section~III and~IV describes the saddle point method and the application of NIG as the base distribution. Sections~V and ~VI describe two specific applications of SPA in determining the outage. Section~V deals with outage due to fading and aggregate interference, while Section~VI describes a stochastic geometric model of downlink COMP transmission. Numerical results are given in Section~VII while Section~VIII concludes the paper.

\textbf{Notations and Definitions:}
$\mathrm{Gamma}(\alpha,\beta)$ denotes Gamma distribution,  $N(a, b)$ represents normal distribution, $IG(a,b)$ denotes the inverse Gaussian distribution, and  $NIG(\alpha,\beta,\mu,\delta)$ denotes  normal-inverse Gaussian distribution, $\mathrm{Poisson}(\lambda)$ denotes Poisson distribution.
$\Gamma(a)=\int_0^\infty x^{a-1} e^{-x} dx$ is the Gamma function, ${\Gamma} (a,b)=\int_b^\infty x^{a-1} e^{-x} dx$ is the upper incomplete Gamma function, and ${\gamma} (a,b)=\int_0^b x^{a-1} e^{-x} dx$ is the lower incomplete Gamma function. $f_X(\cdot)$, $F_X(\cdot)$ and $Q_X(\cdot)$ denote the probability density function (PDF), cumulative density function (CDF), and  complementary CDF of $X$ respectively. The $\Mcal_X(t) = \Ebb[e^{-t X}]$ and $\mu_i(X)$ are the MGF and $i$-th moment of $X$ respectively. The CGF of $X$ is defined as $\Kcal_X(t) = \log \Mcal_X(t)$. The cumulants of $X$, denoted as $\kappa_i(X)$, are the coefficients of Taylor expansion of the CGF, $\Kcal_X(t) = \sum (-1)^i \kappa_i \frac{t^i}{i!}$. The mean, variance, skewness, and excess kurtosis of $X$ are defined in terms of its cumulants as $\Ebb[X] = \kappa_1(X)$, $\Vbb[X] = \kappa_2(X)$, $\Sbb[X] = \frac{\kappa_3(X)}{\kappa_2(X)^{3/2}}$  and $\Kbb[X] = \frac{\kappa_4(X)}{\kappa_2(X)^2}$. In Sections~V and~VI, some of the conventional notations are reused whose interpretation will depend on the local context.

\begin{figure*}
\begin{center}
	\includegraphics[scale=0.2]{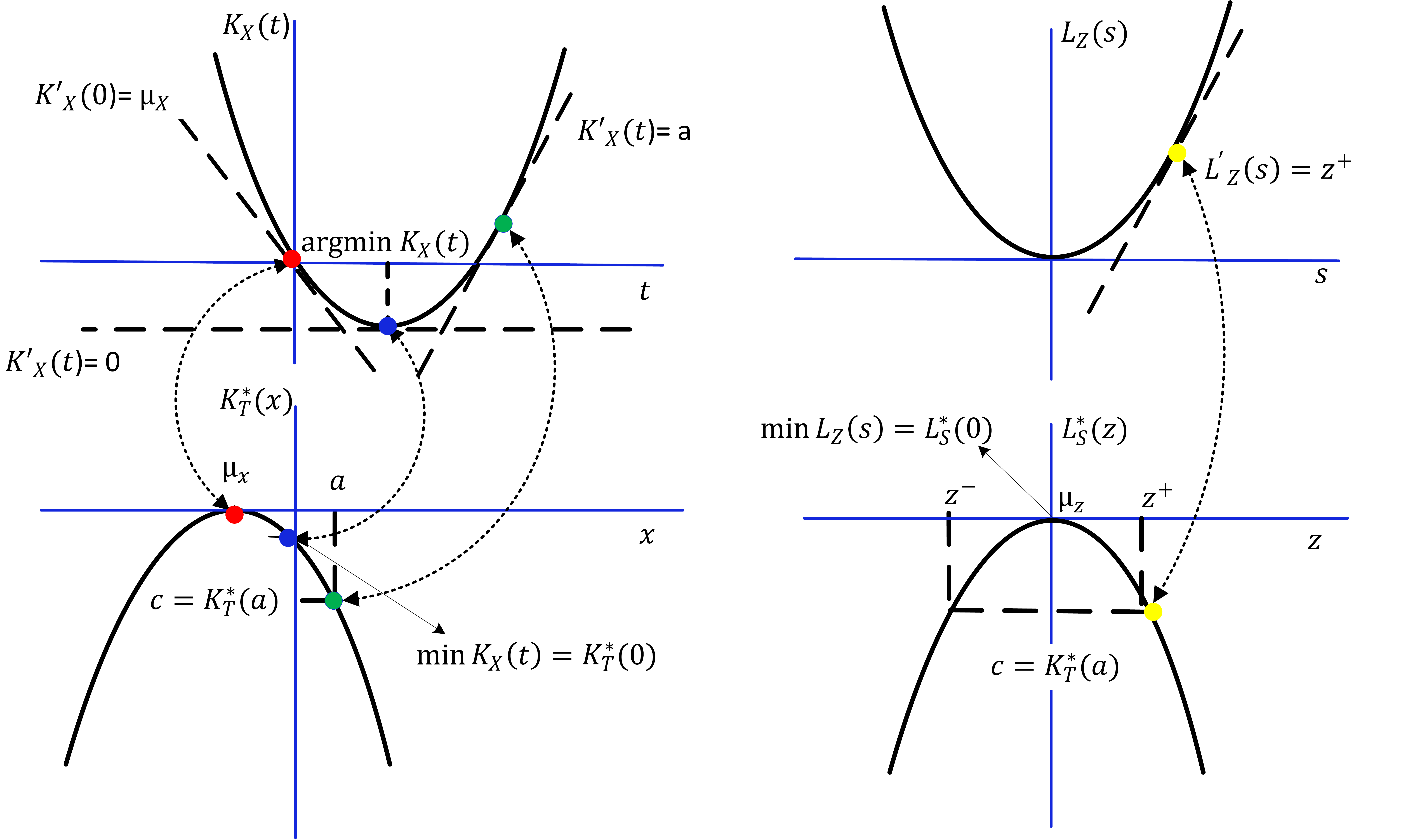}
	\caption{The Legendre-Fenchel duality. }
	\label{fig:legendre-fenchel-duality}
 \end{center}
\end{figure*}

\section{SINR Model and Basic Definitions}

\subsection{SINR Model}
The SINR of a wireless device can be modeled as:
\[ \mathrm{SINR} = \frac{\sum_{i=1}^M X_i}{1 + \sum_{i=1}^N Y_i}=\frac{X}{1+Y},\] 
where $X_i$ denotes the $i$-th useful signal power, $Y_i$ denotes the power of the $i$-th interferer, the noise power is normalized to unity. Here $X = \sum_{i=1}^M X_i$ and $Y =\sum_{i=1}^N Y_i$, and both the $M$ and $N$ can also be a random variables. The variables $X$ and $Y$ may or may not be independent of each other. Following Zhang's approach \cite{Zhang1995}, let us define a new random variable $\Omega = \theta Y - X$, then the SINR outage probability can  be given as follows: 
\beq 
P_{\mathrm{out}} = \mathrm{Pr}(\Omega > -\theta) = Q_\Omega(-\theta).
\label{eqn:outage-def}
\eeq
If we neglect the noise, then the SIR outage will be given by $P_{\mathrm{out}} = \mathrm{Pr}(\Omega > 0)$. 
The SINR outage probability in (\ref{eqn:outage-def}) can be exactly evaluated using Gil-Pelaez inversion formula as~\cite{Gil-Pelaez1951}:
\beq
Q_{\Omega}(\omega) = \frac{1}{2} - \frac{1}{\pi} \int_0^\infty \mathrm{Im}\{\Mcal_\Omega(\jmath t) e^{-\jmath t\omega}\} \frac{\ud t}{t},
\label{eqn:gil-pelaez-outage}
\eeq
where $\mathrm{Im\{z\}}=\frac{z-z^*}{2\jmath}$ is the imaginary component of complex variable $z$, and $\jmath = \sqrt{-1}$.

\subsection{MGF of $\Omega$}
When the signal $X$ and interference $Y$ are correlated, one possible way of obtaining the MGF of $\Omega$ by conditioning on $X$ and then taking an expectation with respect to $X$. Thus we have $\Mcal_{\Omega|X}(t) = \Ebb_Y[e^{-t(\theta Y - X)}|X] = e^{tX} \Mcal_{Y|X}(\theta t)$. Taking the expectation of $\Mcal_{\Omega|X}$ with respect to $X$, we have $\Mcal_{\Omega}(t) = \Ebb_X[e^{tX} \Mcal_{Y|X}(\theta t)]$. In particular, when $X$ and $Y$ are independent of each other, then $\Mcal_{\Omega}(t) = \Mcal_Y(\theta t)\Mcal_X(-t)$. If the signals $X_i$ and $Y_i$ are also mutually independent, then $\Mcal_{\Omega}(t) = \prod_{i=1}^N \Mcal_{Y_i}(\theta t) \prod_{j=1}^M \Mcal_{X_j}(-t)$.

\subsection{CGF and Cumulants of $\Omega$}
Given that $\Mcal_\Omega(\jmath t) = \exp \log{\Mcal_\Omega(\jmath t)} = \exp \Kcal_\Omega(\jmath t)$,  (\ref{eqn:gil-pelaez-outage}) can be restated in terms of CGF as 
\beq
Q_{\Omega}(\omega) = \frac{1}{2} - \frac{1}{\pi} \int_0^\infty \mathrm{Im}\{ e^{\Kcal_\Omega(\jmath t) - \jmath t\omega}\} \frac{\ud t}{t}.
\label{eqn:CGF-omega-generic-case}
\eeq 

Here, the CGF of $\Omega$ conditioned on $X$ is $\Kcal_{\Omega|X}(t) = \log \Mcal_{\Omega|X}(t) = t X + \Kcal_{Y|X}(\theta t)$. Likewise, the CGF of $\Omega$ is $\Kcal_\Omega(t) = \log \Mcal_\Omega(t) = \log \Ebb_X[e^{tX + \Kcal_{Y|X}(\theta t)}]$. In particular, when $X$ and $Y$ are independent, we have the following:

\begin{lemma}[CGF and cumulants of $\Omega$] 
The CGF of $\Omega$ is 
\beq 
\Kcal_\Omega(t) = \log \Ebb_X[e^{tX + \Kcal_{Y|X}(\theta t)}].
\eeq
Furthermore, if $X$ and $Y$ are independent, the CGF of $\Omega$ is 
\beq 
\Kcal_\Omega(t) = \Kcal_Y(\theta t) + \Kcal_X(-t),
\eeq 
and the $n$-th cumulant of $\Omega$ is given by
\beq 
\kappa_n(\Omega) = \theta^n \kappa_n(Y) + (-1)^n \kappa_n(X). 
\label{eqn:nth-cumulant-gamma-generic-model}
\eeq
\label{lemma:CGF-cumulant-omega}
\end{lemma}
\begin{IEEEproof}
Using the additivity and homogeneity properties of CGF and cumulants.
\end{IEEEproof}

If the signals $X_i$ and $Y_i$ are mutually independent, we further have $\Kcal_\Omega(t) = \sum_{i=1}^N \Kcal_{Y_i} (\theta t) + \sum_{j=1}^M \Kcal_{X_j}(-t).$ 

\tbf{Remark:} Independence of $X$ and $Y$ may be invoked in scenarios such as: 1) When the locations of all the BSs in a network are known and the randomness is only due to fading or due to the  number of serving or interfering base stations, 2) When the distance to the serving BS(s) are known and the randomness of the message signal is due only to fading, the locations of interfering base stations being unknown. This is quite a valid assumption in some practical scenarios, especially for location aware services like D2D or for the bipolar networks, 3) The independence assumption can also be invoked for scenarios like the downlink COMP scenario we will study, where independence arises due to the basic property of the underlying Poisson point process. 

 \section{Saddle Point Method}

If the CGF of a random variable $X$ exists, then we can exploit a powerful \tit{saddle point approximation} (SPA)  technique to compute the CDF of the random variable (see \cite{Wong2001, Butler2007} for general introduction). The saddle point method  serves as a compromise between the purely analytical and purely numerical approaches. The direct numerical approach using Gil-Pelaez inversion involves  evaluating the MGF at multiple points to calculate the desired integral. In SPA approach, we only need to evaluate (at least theoretically)  a single point of the function, called the \tit{saddle point}, based on which a semi-analytical formula for outage can be obtained. Also, direct numerical integration of Gil-Peleaz's formula starts to lose its precision and becomes unstable when the value of the integral is extremely small; whereas SPA gives consistent results.

 \subsection{The Wood-Booth-Butler Formula}

Note that the Gil-Pelaez inversion formula can be represented as a contour integral in terms of CGF as 
\[ Q_{X}(x) = \frac{1}{2\pi \jmath} \int_{c- \jmath \infty}^{c+\jmath\infty} e^{\Kcal_X(t) - t x} \frac{\ud t}{t}, \]
where $c>0$ is a real constant lying in the convergence strip of $\Kcal_X(t)$. The dominant component of the integral is concentrated at the neighborhood of the saddle point of $\Kcal_X(t) - tx$. The saddle point $\hat{t} = \hat{t}(x) = \argmin_t\{\Kcal_X(t) - tx\}$ is given by the solution of the saddle point equation 
\beq
\Kcal'_X(\hat{t}) = x.
\label{eqn:key-saddle-pt}
\eeq 
Now, suppose that $g$, $G$, and $\Lcal$ are the PDF, CDF, and CGF of the base distribution of $Z$, respectively, by which we want to approximate our target distribution. The dominant component of this base distribution is found at $\Lcal_Z(\breve{s}) - \breve{s}z$, where $\breve{s} = \breve{s}(z) = \argmin_s\{\Lcal_Z(\breve{s}) - \breve{s}z\}$ is the saddle point root of $\Lcal'_Z(\breve{s}) = z$. After transforming the pair $(x, t) \mapsto (z, s)$ such that dominant components of these two distributions coincide, we obtain
 \beq 
 \Lcal_Z(\breve{s}) - \breve{s} z = \Kcal_X(\hat{t}) - \hat{t}x.
 \label{eqn:coincidence-CGF}
 \eeq
 The task is to find an optimal choice of $\hat{z} = \hat{z}(x)$ from the above transformation process, when the right hand side of (\ref{eqn:coincidence-CGF}) is given. Note that $\Kcal_T^*(x) = \hat{t}x - \Kcal_X(\hat{t})$ and $\Lcal_S^*(z) = \breve{s}z - \Lcal_Z(\breve{s})$ appearing in the left-hand and right-hand of (\ref{eqn:coincidence-CGF}) are Legendre-Fenchel transforms of $\Kcal_X$ and $\Lcal_Z$, respectively, as shown in Fig. \ref{fig:legendre-fenchel-duality}. Succinctly put, (\ref{eqn:coincidence-CGF}) reduces to the matching $\Lcal_S^*(z) = \Kcal_T^*(x)$. The readers are referred to \cite[Ch 3.3]{Borwein2006} for details on Legendre-Fenchel transforms. As per the definition of CGF, both $\Kcal_X(t)$ and $\Lcal_Z(s)$ are convex, whereas their Legendre-Fenchel transforms $\Kcal_T^*(x)$ and $\Lcal_S^*(z)$ are concave with respect to $x$ and $z$. 

In Fig. \ref{fig:legendre-fenchel-duality}, by the property of Legendre-Fenchel transform, the maxima/minima of the dual function is given by the value of the intercept on the ordinate axis of the primal function, whereas the location of the maxima/minima of the dual function is given by the slope of the primal function at that intercept. Thus, $\Kcal_T^*(x)$ and $\Lcal_S^*(z)$ have unique maxima of zero at their means $x = \Ebb[X] = \Kcal'_X(0)$ and $z = \Ebb[Z] = \Lcal'_Z(0)$, respectively. These maxima corresponds to their dual variables $\hat{t}=0$ and $\breve{s} = 0$. Likewise, the minima of $\Kcal_X(t)$ and $\Lcal_Z(s)$ corresponds to the ordinate intercept of their dual function, $\min \Kcal_X(t) = \Kcal_T^*(0)$ and $\min \Lcal_Z(t) = \Lcal_S^*(0)$. 

As such, because of the concavity, for a given value of $x$, there can be two possible optimal choices for $\hat{z}$ in (\ref{eqn:coincidence-CGF}), given by  $\hat{z}_{-}$ and $\hat{z}_{+}$ in Fig. \ref{fig:legendre-fenchel-duality}. For the unique case when $x = \Ebb[X] = \Kcal'_X(0)$, there is only one possible choice of $\hat{z} = \Lcal_Z'(0)$ for the base distribution. For $x \neq \Ebb[X]$, there are two solutions $\hat{z}_{-}(x) < \Ebb[Z] < \hat{z}_{+}(x)$ on either side of the mean for the base distribution. The root of $z$ should be such that its relative position with respect to its mean $\Ebb[Z]$ should  match with the relative position of $x$ with respect to its mean $\Ebb[X]$. Thus,
\beq
\hat{z}(x) = \left\{ \begin{array}{lcr}
\hat{z}_{-}(x), & \mathrm{if} & x < \Ebb[X] \\
\Lcal'_Z(0), & \mathrm{if} & x = \Ebb[X] \\
\hat{z}_{+}(x), & \mathrm{if} & x > \Ebb[X]. 
  \end{array} \right.  
  \label{eqn:optimal-z}
\eeq
Subsequently, we have the following proposition by Wood, Booth, and Butler: 
 
\begin{theorem}[CDF Approximation using SPA \cite{WoodBoothButler1993}]
\label{theo:WBB-formula}
Suppose $X$ has a continuous distribution $F_X(x)$ with CGF $\Kcal_X(t)$. The $(g, G)$-based saddle point CDF approximation for $F_X(x)$ is 
 \beq 
 \hat{F}_X(x) = G_Z(\hat{z}) + g_Z(\hat{z}) \left[ \frac{1}{\hat{s}} - \frac{1}{\hat{u}} \right],
 \label{eqn:WBB-formula}
 \eeq 
 where $\hat{z}$ is given in (\ref{eqn:optimal-z}), $\hat{s} = \breve{s}(\hat{z})$ is the saddle point for $\hat{z}$ with respect to the base CGF and $\hat{u} = \hat{t}\sqrt{\frac{\Kcal''_X(\hat{t})}{\Lcal''_Z(\hat{s})}}$.
 \end{theorem}
 
This CDF approximation is  independent of the location and scale of the base distribution \cite{WoodBoothButler1993}. Also, given a base distribution, the authors recommend the moment matching method to find the parameters of the base distribution, i.e., $ \Lcal_Z^{(n)}(\breve{s}) = \Kcal_X^{(n)}(\hat{t}),$ for $n=1,2,\cdots.$ 

In \tbf{Theorem \ref{theo:WBB-formula}}, $X$ represents a generic random variable whose MGF exists. For our outage problem, since $X \equiv \Omega$, we have an immediately corollary:

\begin{corollary}[SPA of Outage]
\label{coro:SPA-outage}
Given the outage probability as defined in (\ref{eqn:outage-def}), its SPA is 
\beq
P_{\mathrm{out}} = Q_\Omega(-\theta) \simeq 1 - \hat{F}_\Omega(-\theta),
\label{eqn:SPA-outage}
\eeq
where $\hat{F}_\Omega$ is as defined in \tbf{Theorem \ref{theo:WBB-formula}}.
\end{corollary}

\subsection{Lugannani-Rice Formula as a Special Case}

When standard normal distribution is chosen to be the base distribution, we have $\Lcal_Z(s) = \frac{s^2}{2}$. The solution to the saddle point equation $\Lcal'_Z(s) = z$ is simply $\breve{s} = z$, while $\Lcal''_Z(\breve{s}) = 1$. Likewise, solving $\Lcal_Z(\breve{s}) - \breve{s} z = \Kcal_X(\hat{t}) - \hat{t}x$ for $z$ gives $\hat{z} =\sgn(\hat{t})\sqrt{-2(\Kcal_X(\hat{t}) - \hat{t}x)}$. Thus, in the Wood-Booth-Butler formula, we have 
\begin{align}
\hat{s} &= \sgn(\hat{t})\sqrt{-2(\Kcal_X(\hat{t}) - \hat{t}x)}, \\ 
\hat{u} &= \hat{t} \sqrt{\mathcal{K}''_X(\hat{t})}.
\end{align}
This choice of standard normal base distribution leads to the famous \tit{Lugannani-Rice formula} \cite{LugannaniRice1980}. Thus the Lugannani-Rice formula is a special case of Wood-Booth-Butler formula. The Lugannani-Rice formula is extremely robust and very accurate when the distribution to be approximated is nearly Gaussian. However, when the distribution is highly skewed or has very heavy tail, the  results cannot even be considered as probabilities, such as when the result is negative or is greater than unity \cite{BoothWood1994, Annaert2007}. 

 \subsection{Computing the Saddle Point}

Theoretically, the CGF and its derivatives need to be evaluated at just a \tit{single} point  (which is at the saddle point) for SPA; but in practice, this saddle point is not known a-priori. Therefore, the crucial step in applying SPA is in computing the solution of the saddle point equation (\ref{eqn:key-saddle-pt}). Except for a few cases, often the saddle point $\hat{t}$ cannot be found analytically. As such, we have to resort to numerical approaches. Since $\hat{t}$ solves the equation  $\Kcal'_X(\hat{t}) = x$, one possible approach is to apply a root finding algorithm to find $\hat{t}$. For the case when $x=0$,  since $\Kcal_X(t)$ is a convex function, $\hat{t}$ is the global minima. Thus, we can also apply an optimization algorithm to find $\hat{t}$ when $x=0$. 
A proper initialization is the key to minimize the number of iterations  required for these procedures.

For our purpose, an approximation to $\hat{t}$ can be obtained by expanding $\Kcal_X(\hat{t})$ up to second order term, $\Kcal_X(\hat{t}) = - \kappa_1 \hat{t} + \kappa_2 \frac{\hat{t}^2}{2!}$,  and solving for $\Kcal'_X({\hat{t}}) = x$. We have the initial approximation for $\hat{t}$ in terms of first and second central moments as
 \beq
 \hat{t} \approx \frac{x + \kappa_1}{\kappa_2} = \frac{x + \Ebb[X]}{\Vbb[X]}. 
 \label{eqn:initial-approx-SPA}
 \eeq
 This approximation can be used as an initial value for the root finding algorithm or the optimization algorithm. It can also be used to qualitatively understand how $\hat{t}$ adapts according to network parameters. A better approximation (or initialization) of $\hat{t}$ can be obtained by including higher order cumulants in $\Kcal'_X(\hat{t})$ and reverting the series using Lagrange inversion formula \cite{Comtet1970}. When the cumulants are easier to compute than the CGF and its derivatives, as is often the case in stochastic geometry, such reversion of series is recommended to approximate the saddle point.

\section{SPA using NIG distribution}

Now we exploit a  four parameter distribution known as the \tit{normal-inverse Gaussian} (NIG) distribution as a base distribution for SPA framework. NIG distribution is a special case of more general hyperbolic distributions \cite{Nielsen1978}. NIG distribution allows us great flexibility in adjusting the shape of the  distribution and the decay rates of the tail. The NIG distribution was first introduced in \cite{Eberlein1995,Nielsen1997} to model financial processes and has since then found many applications. It was introduced to signal processing in \cite{Hanssen2001}. 
Since the NIG distribution is defined on the entire real line, it is suitable for the modeling of random variable $\Omega$. 
 
NIG distribution can be defined as follows: Suppose $X$ is normal distributed when conditioned on $Y$, with mean $\mu + \beta Y$ and variance $Y$, i.e., $f_{X|Y}(x|y) = N(\mu+\beta Y, Y)$. Now if $Y$ itself follows an inverse Gaussian distribution $f_Y(y) = IG(\delta,\sqrt{\alpha^2 - \beta^2})$, then the unconditional distribution of $X$ is said to be normal-inverse Gaussian $f_X(x) = NIG(\alpha,\beta,\mu,\delta)$. The PDF of NIG distribution is given by 
 \begin{align}
 f_Z(z; \alpha, \beta, \mu, \delta) &=  \frac{\alpha}{\pi \delta}\exp(\delta \gamma + \beta(z-\mu)) \times \frac{K_1 \left(\alpha \delta \sqrt{1+(\frac{z-\mu}{\delta})^2}\right)}{\sqrt{1+(\frac{z-\mu}{\delta})^2}},
 \label{pdf-nig}
 \end{align}
where $z\in\Rbb$, $\alpha>0$, $\delta>0$, $\mu\in\Rbb$, $0<|\beta|<\alpha$, and $\gamma = \sqrt{\alpha^2 - \beta^2}$. The $K_\nu(\cdot)$ is modified Bessel function of second kind\footnote{In some texts, $K_\nu$ is referred to as modified Bessel function of third kind.} 
 \[ K_\nu(z) = \frac{1}{2} \int_0^{\infty} u^{\nu -1} \exp \left({-\frac{1}{2}z(u+u^{-1})} \right) \ud u, \]
 with index $\nu = 1$. The $\mu$ represents location parameter, $\alpha$ represents tail heaviness, $\beta$ represents asymmetry parameter, $\delta$ represents scale parameter.  
 
Large $\alpha$ implies light tails, while smaller $\alpha$ implies heavier tails. Similarly, $\beta < 0$ implies left skewness, $\beta > 0$ implies right skewness, while $\beta = 0$ implies that the distribution is symmetric. Furthermore, the symmetric NIG distribution tends to Gaussian distribution as $\alpha \rightarrow \infty$ and $\delta \propto \alpha$. On the other hand, the symmetric NIG distribution tends to Cauchy distribution as $\alpha \rightarrow 0$. 
 In Wolfram Mathematica~10.0, the NIG distribution is available as a  built-in function {\texttt{HyperbolicDistribution}[$-1/2,\alpha,\beta,\delta,\mu$]}. 

The mean, variance, skewness, and excess kurtosis of NIG distribution are:
 \[ \begin{array}{ll}
\Ebb[Z] = \mu + \frac{\delta \beta}{\gamma},  &
\Vbb[Z] = \frac{\delta \alpha^2}{\gamma^3}, \\
\Sbb[Z] = \frac{3}{\sqrt{\delta \gamma}}\left(\frac{\beta}{\alpha}\right), & 
\Kbb[Z] = \frac{3}{\delta \gamma}\left( 1 + 4 \left( \frac{\beta}{\alpha} \right)^2\right).
 \end{array}
 \]
 The parameters of the distribution can be explicitly solved, given the first four cumulants, using the moment matching method \cite{Eriksson2004}. It is easy to see that the skewness is bounded by excess kurtosis as 
 \beq
 \Kbb[Z] \geq \frac{4}{3} \Sbb^2[Z].
 \label{eqn:kurt-skew-NIG-bound}
 \eeq 
Any variable that satisfies this inequality can thus be modeled by NIG distribution. 
 
Despite the appearance of modified Bessel function in the definition of PDF of NIG distribution, the CGF of NIG distribution has a much simpler form: 
 \beq
 \Lcal_Z(s) = \mu s + \delta [\sqrt{\alpha^2 - \beta^2} - \sqrt{\alpha^2 - (\beta + s)^2}].
 \eeq
The algebraic simplicity of its CGF makes NIG distribution a good candidate for the base distribution for SPA, since the saddle point of NIG distribution can be analytically solved.

\begin{theorem}
 Let $\eta = \frac{\Kcal'''(\hat{t})^2}{\Kcal''(\hat{t})^3}$, $\rho = \frac{\Kcal^{\textrm{\romannumeral 4}}(\hat{t})}{\Kcal''(\hat{t})^2}$, and $c = \Kcal_X(\hat{t}) - x\hat{t}$. Let the location and scaling parameters of NIG distribution be selected as $\mu=0$ and $\delta = 1$. Assuming $c<0$ and $0 \leq \rho - \frac{5}{3} \eta \leq \frac{3}{|c|}$, the parameters for Wood-Booth-Butler formula are:
 \begin{enumerate}
\item The saddle point $\breve{s}$ of NIG base distribution for given $z$ is
 \beq  \breve{s}(z) = -\beta + \frac{\alpha z}{\sqrt{1+ z^2}}. \eeq
 \item The optimal choice of $\hat{z}$ for given $x$ and $\hat{t}$ is 
 \beq  \hat{z} = \sgn(\hat{t})\left(\frac{3\rho}{\eta} - 5\right)^{-1/2}. \eeq
 \item The second derivative of $\Lcal_Z$ at $\breve{s}$ given $z$ is 
 \beq L''_Z(\breve{s}) = \frac{z^3 + z}{\breve{s}+\beta}. \eeq
 \item Lastly, the $\alpha$ and $\beta$ parameters of the base NIG distribution, when $\mu =0$ and $\delta = 1$, are
 \begin{align}
 \alpha &= 9 [(3\rho - 5 \eta)(3\rho - 4\eta)]^{-1/2}, \\
 \beta &= \frac{e \hat{z} + \sgn(\Kcal_X'''(t)) \sqrt{\alpha^2(1 + \hat{z}^2) -e^2}}{1+\hat{z}^2},
 \end{align}
where $e = c + \alpha\sqrt{1+\hat{z}^2}$.
 \end{enumerate}
 \label{theorem:SPA-NIG}
\end{theorem} 
  
\begin{IEEEproof}
See \tbf{Appendix~A}. 
\end{IEEEproof}


The sufficiency condition that appears in \textbf{Theorem \ref{theorem:SPA-NIG}} is not surprising given (\ref{eqn:kurt-skew-NIG-bound}). However, the sufficiency conditions are not always guaranteed to be satisfied. In such cases, we can use the symmetric NIG distribution as the base distribution.

\begin{theorem}
Let $\rho = \frac{\Kcal^{\textrm{\romannumeral 4}}(\hat{t})}{\Kcal''(\hat{t})^2}$, and $c = \Kcal_X(\hat{t}) - x\hat{t}$. Let the location and asymmetry parameter of NIG distribution be selected as $\mu=0$ and $\beta=0$. Assuming that a solution $v$ exists for the cubic equation 
\beq 
5 v^3 - 5v^2 + \left(\frac{\rho c}{3}-4\right) v + 4 = 0,
\label{eqn:cubic-cond-sym-NIG}
 \eeq
such that $v < -1$ if $c<0$ and $v > 1$ if $c >0$, then the parameters for Wood-Booth-Butler formula are:
 \begin{enumerate}
 \item The saddle point $\breve{s}$ of NIG base distribution for given $z$ is
 \beq  \breve{s}(z) = -\beta + \frac{\alpha z}{\sqrt{\delta^2+ z^2}}. \eeq
 \item The optimal choice of $\hat{z}$ for given $x$ and $\hat{t}$ is 
 \beq  \hat{z} = \sgn(\hat{t})\sqrt{c(v-1)}. \eeq
 \item The second derivative of $\Lcal_Z$ at $\breve{s}$ given $z$ is 
 \beq L''_Z(\breve{s}) = \frac{\delta}{\breve{s}+\beta}\left[\left(\frac{z}{\delta}\right)^3 + \left(\frac{z}{\delta}\right)\right] . \eeq
 \item Lastly, the $\alpha$ and $\delta$ parameters of the base NIG distribution, when $\mu =0$ and $\beta = 1$, are
 \beq \alpha = \delta = \sqrt{\frac{c}{1+v}}.\eeq
 \end{enumerate}
  \label{theorem:SPA-symmetric-NIG}
\end{theorem}

\begin{IEEEproof}
See \tbf{Appendix~B}.
\end{IEEEproof}

\tbf{Remark:} For the SIR outage approximation, we evaluate the saddle point at $\omega = 0$. This point corresponds to the minima of the CGF of $\Omega$ such that $\Kcal_\Omega(\hat{t}) < 0$. Thus the condition $c<0$ is always satisfied.

\section{Application of SPA for IID Random Variables}

Consider the special case where $\Omega = \theta Y - X$, $Y = \sum_{i=1}^N G_i$, and $X = \sum_{j=1}^M G_j$ such that $\{G_i\}$ are independent and identically distributed (IID) random variables. Also, let the variables $M$ and $N$ be random. The CGFs of compound distribution is $\Kcal_Y(t) = \Kcal_N(-\Kcal_G(t))$ and $\Kcal_X(t) = \Kcal_M(-\Kcal_G(t))$, while $\Kcal_\Omega(t) = \Kcal_Y(\theta t) + \Kcal_X(-t)$.   In the following, we will consider the cases when $N$ and $M$ obey Poisson and Binomial distributions, and when $G_i$ follows some simple distribution. After finding the saddle point, the outage can be computed using \tbf{Corollary} \ref{coro:SPA-outage} of \tbf{Theorem} \ref{theo:WBB-formula}.

The simplicity of this setting is worthy of theoretical examination for its own sake, since it presents one of those few cases where the saddle point can be computed analytically. One such instance where this scenario can arise is in a multi-user setting where a BS transmits to a user, and the user performs receive diversity combining, while experiencing co-channel interference from the signals intended for other users. Thus, the randomness will arise  due to fading and/or number of interferers.

\subsection{Nakagami-$m$ Fading and Poisson Aggregation}

For Nakagami-$m$ fading, the channel power gain is given by the Gamma distribution, $G_i \sim \mathrm{Gamma}(\alpha,\beta)$, such that its MGF is $\Mcal_G(t) = (1+ \frac{t}{\beta})^{-\alpha}$. Let the Poisson aggregation be given by $M \sim \mathrm{Poisson}(\lambda_1)$ and $N \sim \mathrm{Poisson}(\lambda_2)$. Using the relation for compound Poisson distribution, we have 
$\Kcal_Y(t) = \lambda_2[(1+\frac{t}{\beta})^{-\alpha} - 1]$ and $\Kcal_X(t) = \lambda_1[(1+\frac{t}{\beta})^{-\alpha} - 1]$. Thus, we have from (\ref{eqn:CGF-omega-generic-case})
\begin{align}
\Kcal_\Omega(t) &= \lambda_2 \left[\left(1+\frac{\theta t}{\beta}\right)^{-\alpha} - 1\right] + \lambda_1 \left[\left(1 - \frac{t}{\beta}\right)^{-\alpha} - 1\right].
\end{align}
We can obtain the saddle point for this case analytically.

\begin{proposition}
Let $\zeta = (\frac{\theta \lambda_2}{\lambda_1})^{-\frac{1}{\alpha + 1}}$. For Nakagami-$m$ fading and Poisson aggregation, the saddle point of $\Kcal_\Omega(t)$ is 
\beq 
\hat{t} = \beta \frac{1 - \zeta}{1 + \theta \zeta}. 
\label{eqn:saddle-pt-poisson-nakagami}
\eeq
\end{proposition}

\begin{IEEEproof}
The derivative of $\Kcal_\Omega(t)$ with respect to $t$ is 
\[ \Kcal'_\Omega (t) = -\frac{\alpha  \theta  \lambda _2}{\beta} \left(1+\frac{\theta  t}{\beta }\right)^{-\alpha -1} + \frac{\alpha  \lambda _1}{\beta } \left(1-\frac{t}{\beta }\right)^{-\alpha -1}. \]
Solving the saddle point equation $\Kcal'_\Omega(\hat{t}) = 0$ for $\hat{t}$, after basic algebra, we have the desired result.
\end{IEEEproof}

%
%


\tbf{Remark:} The parameters $\lambda_1$ and $\lambda_2$ can be interpreted as  a thinning of a parent PPP with parameter $\lambda$ over a common spatial area, such that $\lambda_1 = p \lambda_1$ and $\lambda_2 = (1 - p) \lambda_2$, where $p$ can be interpreted as probability of cooperation. 
Alternatively, $\lambda_1$ and $\lambda_2$ can arise due to Poisson point process over two mutually exclusive spatial regions of differing areal sizes, as in our considered cellular network model. In particular, when there is no interferers $\lambda_2 = 0$, then $\hat{t} = \beta$.

%
%
 \subsection{Nakagami-$m$ Fading and Binomial Aggregation}

  Now, consider instead the case when we have $L$ total nodes such that $M \sim \mathrm{Binomial}(L,p)$ and $N = L - M \sim \mathrm{Binomial}(L,q)$, where $p+q = 1$. Here $p$ is interpreted as the probability of cooperation. As before, for Nakagami-$m$ fading, the channel gain is given by the gamma distribution, $G_i \sim \mathrm{Gamma}(\alpha, \beta)$, such that its MGF is $\Mcal_G(t) = (1+ \frac{t}{\beta})^{-\alpha}$. Using the relation for compound binomial distribution, we have $\Kcal_Y(t) = L \log(p + q (1 + \frac{t}{\beta})^{-\alpha})$ and $\Kcal_X(t) = L \log(q + p (1 + \frac{t}{\beta})^{-\alpha})$. Hence, we have from (\ref{eqn:CGF-omega-generic-case}) 
 \begin{align}
 \Kcal_\Omega(t) &=  L \log \left[\left(p + \frac{q}{ \left(1 + \frac{\theta t}{\beta}\right)^{\alpha}}\right) \left(q + \frac{p}{\left(1 - \frac{t}{\beta}\right)^{\alpha}} \right)\right].
 \end{align}
 
 \begin{proposition}
 For Nakagami-$m$ fading and binomial aggregation, the solution to $\Kcal'_\Omega(\hat{t}) = 0$ is found by solving 
 \[ p^2\left(1+\frac{\theta \hat{t}}{\beta}\right)^{\alpha+1} + pq\left(1+\theta\right)\frac{\hat{t}}{\beta} - q^2\left(1-\frac{\hat{t}}{\beta}\right)^{\alpha+1} = 0. \] 
  \end{proposition}
 \begin{IEEEproof}
 Taking the derivative of $\Kcal_\Omega(t)$ with respect to $t$, the saddle point equation $\Kcal'_\Omega(\hat{t}) = 0$ can be simplified to obtain the desired result.
  \end{IEEEproof}
  
  
For the special case when there are no interferers $q=0$, so $\hat{t} = -\beta/\theta$. Unfortunately, we cannot in general solve the saddle point equation analytically, and we need to resort to numerical root finding technique. However, for Rayleigh fading case, we have: 
 \begin{corollary}
 For the Rayleigh fading, where $\alpha = 1$, 
 \beq 
 \hat{t} = \frac{- 2 \beta \theta (1 - pq) + 2 \beta \sqrt{pq\theta(\theta+q)(1+p\theta)}}{2 \theta (\theta p^2 - q^2) }.
 \label{eqn:saddle-pt-binomial-rayleigh}
 \eeq
  \end{corollary}
 
 Nevertheless, we can obtain an approximation for the saddle point for the general case as given in (\ref{eqn:initial-approx-SPA}) as
 \beq
\hat{t} \approx \frac{\beta  (p-\theta  q)}{1 - q \left[1 - \theta ^2 - \alpha  \left(\theta ^2+1\right) (1-q)\right]}.
 \eeq
 Here, we see that the saddle point is independent of $L$. Also, for $q\rightarrow 0$, $\hat{t} \approx \beta$; and for $q\rightarrow 1$, $\hat{t} \approx -\beta/\theta$. Similarly, for $\theta \rightarrow 0$, $\hat{t} \approx \beta p/ (1-q(1-\alpha p))$; and when $\theta \rightarrow \infty$, $\hat{t} \approx - \beta/(\theta(1+\alpha pq))$.

\section{Application of SPA in Downlink COMP Transmission}

In this section, we will confine ourselves to a simple stochastic geometrical model (see \cite{Haenggi2012} for general introduction) for COMP. More sophisticated models have been investigated in \cite{Nigam2014} -- \cite{Tanbourgi2014}. The purpose here is not to investigate COMP for its own sake, but rather to illustrate the use and compare the accuracy of different SPA methods.

\subsection{Spatial Cellular Network Model} 

\begin{figure}[h]
\begin{center}
	\includegraphics[width=3in]{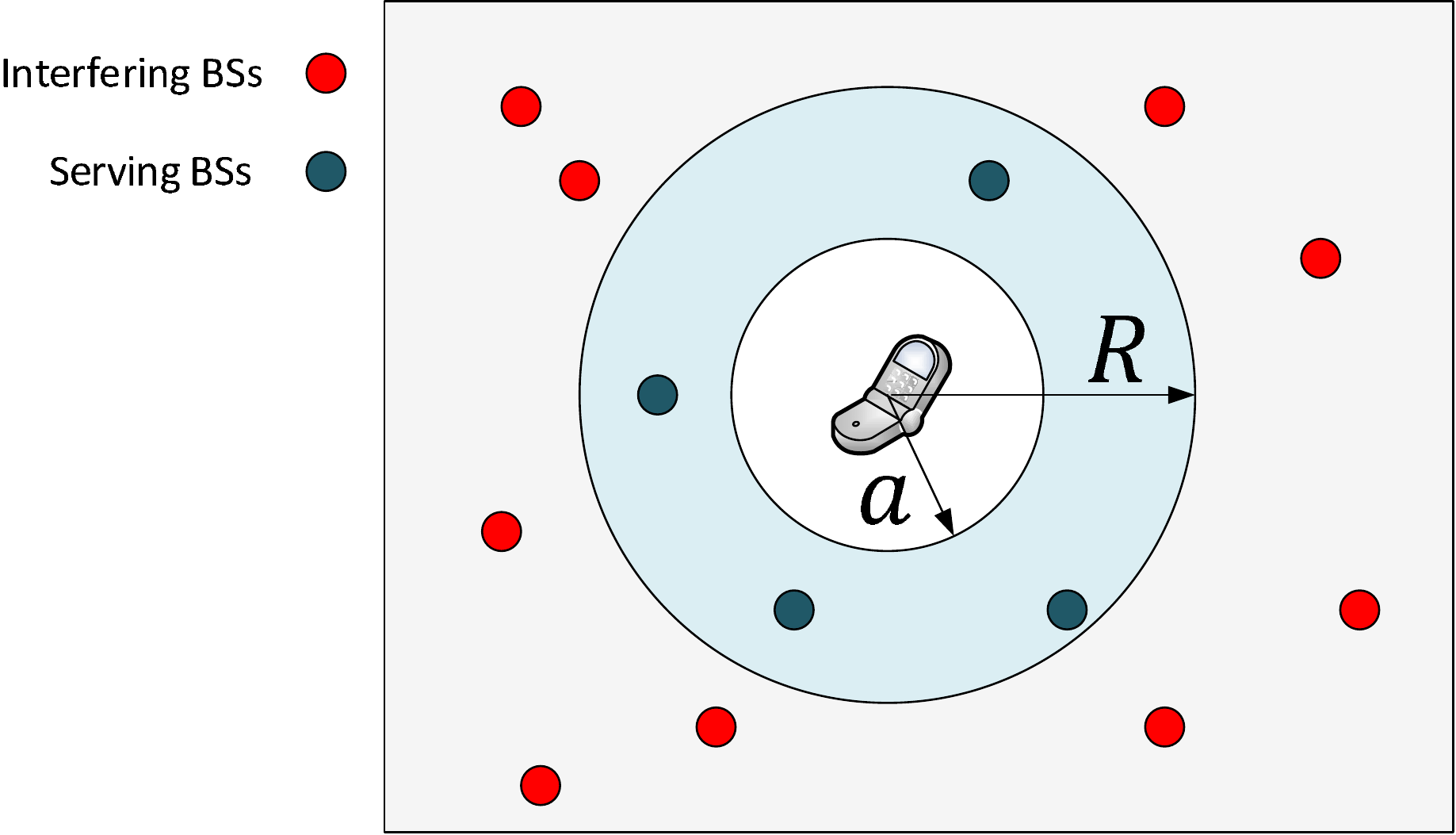}
	\caption{The exclusion region, the region of cooperation, and the region of interference. }
	\label{fig:different-zones}
 \end{center}
\end{figure}

Let single antenna BSs be scattered in 2-D plane according to a homogeneous PPP of intensity $\lambda$. Consider an annular region $\Bcal^c$ centered at origin and with \tit{fixed} outer radius $R$ and inner radius $a > 0$, such that $\Bcal^c = \{r | a \leq r < R\}$. Consider a typical single-antenna user equipment (UE) located at the origin, as shown in Fig. \ref{fig:different-zones}. It is assumed that there are no BSs located within radius $r < a$, thus forming an \tit{exclusion region} for the typical UE. All BSs within $\Bcal^c$  cooperate with each other to conduct a coordinated multi-point (COMP) transmission to the typical UE. All BSs beyond $R$,  $\Bcal^{nc} = \{r| r \geq R\}$, act as interferers; thus $\Bcal^{nc}$ forms the \tit{interference region}. Since $\Bcal^{c}\cap \Bcal^{nc} = \varnothing$, the BSs in $\Bcal^{c}$ and $\Bcal^{nc}$ are both PPP of intensity $\lambda$, as per the property of PPP\footnote{The same scenario can also be extended for the uplink when  the typical UE transmits a message, which is then cooperatively detected by the BSs within $\Bcal^c$.}. 

Note that, $R$ is a design parameter which can be selected according to the need. For instance, we can design the disk radius $R$ such that the \tit{average} number of BSs inside the disk radius is $k$, giving an approximation of the COMP comprising of $k$-nearest BSs.

\subsection{Received Signal and Interference Model} 

The instantaneous received signal of the typical UE is modeled as
$v = \sum_{i\in\Phi_{\Bcal^c}} \sqrt{g_i} r_i^{-\frac{\alpha}{2}} s + \sum_{j\in\Phi_{\Bcal^{nc}}} \sqrt{g_j} r_j^{-\frac{\alpha}{2}} s_j + z, $
 where $s$ is the common message signal transmitted by all BSs in $\Bcal^{c}$ and $s_j$ are interfering signals from BSs in $\Bcal^{nc}$, $r_i$ is the distance between $i$-th BS and the typical UE, $g_i$ denotes independent and identically distributed (IID) channel gains, and $\alpha$ is the path-loss exponent {such that $\alpha > 2$}. Lastly, $z$ is the additive white noise. 
Let the variances $\Vbb(s) = \Vbb(s_i) = P$ and $\Vbb(z) = \sigma^2$. Assuming maximum-ratio-combining (MRC) at the typical UE, we can write the SINR as follows:
\begin{equation}\label{sinr}
\mathrm{SINR} = \frac{\sum_{i\in\Phi_{\Bcal^{c}}} g_i P r_i^{-\alpha}}{\sum_{j\in\Phi_{\Bcal^{nc}}} g_j P r_j^{-\alpha} + \sigma^2}.
\end{equation}
Defining $X = \sum_{i\in\Phi_{\Bcal^{c}}} g_i P r_i^{-\alpha}$, and $Y = \sum_{j\in\Phi_{\Bcal^{nc}}} g_j P r_j^{-\alpha}$, we have SINR outage probability \[P_{\mathrm{out}} = \mathrm{Pr}(\Omega > -\theta \sigma^2),\] where $\Omega = \theta Y - X$ and $\theta$ is the predefined SINR threshold. For simplicity, we will assume the noise to be zero. Thus, we will consider SIR outage, rather than SINR outage. Lastly, the SPA for the outage is given by \tbf{Corollary \ref{coro:SPA-outage}} of \tbf{Theorem \ref{theo:WBB-formula}}.

\subsection{CGF of $\Omega$ for the COMP Model}

\begin{lemma}
The CGF of $\Omega$ for the considered large scale cellular network model is given as
\begin{align} 
\Kcal_\Omega(t) 
&= 2 \pi \lambda \big[ \int_a^R (\Mcal_G(-t P r^{-\alpha}) - 1) r \ud r  + 
  \int_R^\infty (\Mcal_G(t \theta P r^{-\alpha}) - 1) r \ud r  \big]. 
\label{eqn:CGF-omega-rep-case}
\end{align}
\label{prop:CGF-omega-rep-case}
\end{lemma}
\begin{IEEEproof}
For homogeneous PPP, the CGF of aggregate of impulse response positioned at each 2-D Poisson point is {$\Kcal(t) = 2 \pi \lambda \int (\Ebb_G[e^{-t g \ell(r)}] - 1) r \ud r$}, where $\ell(r)$ is the deterministic path-loss function and $g$ is the random channel gain. For our case, $\ell(r) = Pr^{-\alpha}$. Thus, $\Kcal_X(-t) = 2 \pi \lambda  \int_a^R (\Ebb_G[e^{t g P r^{-\alpha}}] - 1) r \ud r ]$ and $\Kcal_Y(\theta t) = 2 \pi \lambda \int_R^\infty (\Ebb_G[e^{- t \theta g P r^{-\alpha}}] - 1) r \ud r ] $. Note that $\Ebb_G [e^{t g P r^{-\alpha}}] \equiv \Mcal_G(-t P r^{-\alpha})$ and $\Ebb_G [e^{-t \theta g P r^{-\alpha}}] \equiv \Mcal_G (t \theta P r^{-\alpha})$. Consequently, using (\ref{eqn:CGF-omega-generic-case}) of \textbf{Lemma~\ref{lemma:CGF-cumulant-omega}} we have the desired result.
\end{IEEEproof}

Here the non-zero lower limit $a$ allows us to avoid the singularity at origin of the unbounded path-loss function $\ell(r)$. Also, it allows us to model the exclusion region. It is very important to set this parameter correctly, since it determines the heaviness of the tail of the resulting distributions. Small exclusion regions produce distributions with heavier tails while large exclusion regions produce distributions with lighter tails. 

%
%
 \subsection{Cumulants of $\Omega$ for the COMP Model} 
 
 \begin{lemma}
 The $n$-th cumulant of $\Omega$ for our large scale cellular network model is given by
\beq 
\kappa_n(\Omega) = \kappa_n^{\mathrm{lim}} (\Omega) [1 + ((-\theta)^n - 1) u^{-n\alpha + 2}], 
\eeq
where $u = R/a$ and $ \kappa_n^{\mathrm{lim}} (\Omega) = ( -1)^n \frac{2 \pi  \lambda \mu_n(G) P^n}{n\alpha-2} a^{-n\alpha+2}$.
\label{prop:cumulant-rep-model}
 \end{lemma}
 
 \begin{IEEEproof}
Using the Campbell's formula for signal $X$, evaluating the integral over the limits $a$ and $R$, we have 
\begin{eqnarray}
\kappa_n(X) &  =  & 2 \pi \lambda \mu_n(G) P^n \int_a^R r^{-n\alpha + 1} \ud r 
= \frac{2 \pi \lambda \mu_n(G) P^n}{n\alpha-2}(a^{-n\alpha+2} - R^{-n\alpha+2}) . 
\end{eqnarray}
Similarly, for interference $Y$, the limits of integral are $R$ to $\infty$; therefore,
\begin{eqnarray}
\kappa_n(Y) & = & 2 \pi \lambda \mu_n(G) P^n \int_R^\infty r^{-n\alpha + 1} \ud r 
     = \frac{2 \pi \lambda \mu_n(G) P^n}{n\alpha-2} R^{-n\alpha+2}. 
\end{eqnarray}
Substituting the expressions for $\kappa_n(X)$ and $\kappa_n(Y)$ in (\ref{eqn:nth-cumulant-gamma-generic-model}) of \textbf{Lemma~\ref{lemma:CGF-cumulant-omega}}, we obtain 
\begin{eqnarray}
\kappa_n(\Omega) & = & \frac{2 \pi \lambda \mu_n(G) P^n}{n\alpha-2} a^{-n\alpha+2} 
  \left[ \theta^n \Big( \frac{R}{a} \Big)^{-n\alpha + 2} + ( -1)^n \Big\{ 1 - \Big( \frac{R}{a} \Big)^{-n\alpha + 2} \Big\} \right].  \nonumber \\
\end{eqnarray}
Simplifying, we obtain the desired result.
\end{IEEEproof}
 
 \begin{corollary} $\lim_{u \rightarrow \infty} \kappa_n(\Omega) = \kappa_n^{\mathrm{lim}}(\Omega). $ 
 \label{coro:lim-cumulant-rep-model}
 \end{corollary}
 
 \begin{corollary} Assuming $a<<R$, so that $\kappa_n(\Omega) \sim \kappa_n^{\mathrm{lim}}(\Omega)$, the skewness squared and excess kurtosis of $\Omega$ are
 \begin{align*}
 \Sbb^2[\Omega] &= \frac{1}{2\pi\lambda a^2}\frac{(\alpha-1)^3}{(3\alpha-2)^2} \Sbb^2[G], \qquad
 \Kbb[\Omega] = \frac{1}{\pi\lambda a^2}\frac{(\alpha-1)^2}{2\alpha-1} \Kbb[G].
 \end{align*}
 
 \label{coro:skew-kurtosis-rep-model}
 \end{corollary}
 
 \tbf{Remark:} According to \textbf{Corollary \ref{coro:lim-cumulant-rep-model}}, when $a << R$, the cumulants are independent of threshold $\theta$. Also, according to \textbf{Corollary \ref{coro:skew-kurtosis-rep-model}}, without losing much generality, we can see that both the skewness and kurtosis of $\Omega$ decreases as $a$ and $\lambda$ increases. This implies that the distribution of $\Omega$ is approximately Gaussian only for large $a$ and $\lambda$. In the context of SPA, this further implies that we can reliably use Lugannani-Rice formula only for large $a$ and $\lambda$.

 \subsection{Approximate Saddle Point}
 
 Using the cumulants of $\Omega$, we can now have an initial approximation of the saddle point $\hat{t}$ for SPA, evaluated at $\omega = 0$, from (\ref{eqn:initial-approx-SPA}) as 
 \begin{align*}
 \hat{t} &\approx -\frac{2(\alpha-1)}{\alpha - 2} \frac{\mu_1(G)}{\mu_2(G)} \left(\frac{a^\alpha}{P}\right) \left(\frac{1-(1+\theta)u^{-\alpha+2}}{1-(1-\theta^2)u^{-2\alpha+2}}\right).
 \end{align*}
A surprising aspect of this approximation is that the saddle point does not depend on $\lambda$. If we further consider the case when the threshold SINR is very high, $\theta \rightarrow \infty$, then we have further simplification of the last term
 \begin{align*}
 \frac{1-(1+\theta)u^{-\alpha+2}}{1-(1-\theta^2)u^{-2\alpha+2}} &\approx  - \frac{\theta u^{-\alpha+2}}{\theta^2 u^{-2\alpha+2}} 
 = \frac{-1}{\theta}\left(\frac{R}{a}\right)^\alpha.
 \end{align*}
 This yields the approximate saddle point as
 \begin{align}
 \hat{t} \approx \frac{2(\alpha-1)}{\alpha - 2} \frac{\mu_1(G)}{\mu_2(G)} \left( \frac{R^\alpha}{\theta P}\right).
 \label{eqn:sp-approx-case-C-large-threshold}
 \end{align}
 Thus we see that the saddle point is unaffected by changes in $a$ and $\lambda$ when $\theta$ is large. Similarly, when the threshold SINR is small, $\theta \rightarrow 0$, then we have
 \[\frac{1-(1+\theta)u^{-\alpha+2}}{1-(1-\theta^2)u^{-2\alpha+2}} \approx  \frac{1 - u^{-\alpha+2}}{1 - u^{-2\alpha+2}} \approx 1. \]
Thus we have the approximate saddle point as 
\beq 
\hat{t} \approx \frac{2(\alpha-1)}{\alpha - 2} \frac{\mu_1(G)}{\mu_2(G)} \left( \frac{a^\alpha}{P}\right),  
\label{eqn:sp-approx-case-C-small-threshold}
\eeq
 which is unaffected by changes in $R$ and $\lambda$.

 \subsection{Special Case: When Fading is Absent}
 
 From our representative model, we have the CGF of $\Omega$ as given by (\ref{eqn:CGF-omega-rep-case}). If fading is absent, then the channel is  deterministic. Thus the CGF of $\Omega$ simplifies to 
 \beq 
 \Kcal_\Omega(t) = 2 \pi \lambda \left[ \int_a^R (e^{t P r^{-\alpha}} - 1) r \ud r + \int_R^\infty (e^{- t \theta P r^{-\alpha}} - 1) r \ud r \right], 
 \label{eqn:CGF-omega-rep-case-C}
 \eeq
 where the channel gain is normalized to unity. The integrals and the derivatives of (\ref{eqn:CGF-omega-rep-case-C}) can be evaluated using generalized incomplete Gamma function defined as $\Gamma(a,z_0,z_1) = \Gamma(a,z_0)-\Gamma(a,z_1)$, where $\Gamma(a,z)=\int_z^\infty x^{a-1}e^{-x}\ud x$ is the upper incomplete Gamma function. To find the derivatives of $\Kcal_\Omega(t)$, we will first give the following proposition.
 
\begin{proposition}
If $\Kcal(t) = 2 \pi \lambda \int_a^b (e^{-t P r^{-\alpha}} - 1) r \ud r$, then its $n$-th derivative is
\begin{align} 
\Kcal^{(n)}(t)  &=  (-1)^n \frac{2\pi\lambda}{\alpha} \frac{(tP)^{2/\alpha}}{t^n}  \Gamma\left(-\frac{2}{\alpha},tPb^{-\alpha},tPa^{-\alpha}\right). 
\label{eqn:n-th-derivative-CGF-rep-case-C}
\end{align}
\label{prop:n-th-derivative-CGF-rep-case-C}
\end{proposition}

\begin{IEEEproof}
See \textbf{Appendix~C}.
 \end{IEEEproof}

Since we have $\Kcal_\Omega(t) = \Kcal_Y(\theta t) + \Kcal_X(-t)$, the derivatives of $\Kcal_\Omega(t)$ immediately follow by applying (\ref{eqn:n-th-derivative-CGF-rep-case-C}).
\begin{proposition}
The $n$-th derivative of $\Kcal_\Omega(t)$ is $\Kcal^{(n)}_\Omega(t) = \Kcal^{(n)}_Y(\theta t) + \Kcal^{(n)}(-t)$, where 
\begin{align*}
\Kcal^{(n)}_Y(\theta t) &= (-1)^n \frac{2\pi\lambda}{\alpha} \frac{(\theta tP)^{2/\alpha}}{t^n} \gamma\left(-\frac{2}{\alpha}, t\theta PR^{-\alpha}\right), \\
\Kcal^{(n)}_X(-t) &= (-1)^n \frac{2\pi\lambda}{\alpha} \frac{(-tP)^{2/\alpha}}{t^n} \Gamma\left(-\frac{2}{\alpha}, -tPR^{-\alpha},  -tPa^{-\alpha}\right), 
\end{align*}
where $\gamma(a,z)$ is the lower incomplete Gamma function, such that $\gamma(a,z)+\Gamma(a,z) = \Gamma(a)$.
\end{proposition}

\begin{IEEEproof}
By applying (\ref{eqn:n-th-derivative-CGF-rep-case-C}) of Proposition \ref{prop:n-th-derivative-CGF-rep-case-C} to $\Kcal_Y(\theta t)$ and $\Kcal_X(-t)$.
\end{IEEEproof}

Since $\Gamma(a,-z)$ is in general a complex number, we have to be careful when interpreting this result. To solve the saddle point equation $\Kcal'_\Omega(t) = 0$, we need to resort to numerical root finding techniques. The approximations for the saddle point given in (\ref{eqn:sp-approx-case-C-large-threshold}) and (\ref{eqn:sp-approx-case-C-small-threshold}) also applies here, but without the moments of $G$.

\section{Numerical Results}

In this section, we will present numerical results that compare the outage probability obtained via direct numerical integration of Gil-Pelaez formula and SPA as given by the Wood-Booth-Butler formula. For simplicity, we assume the noise to be zero, thus obtaining SIR outage, rather than SINR outage. For the saddle point method, we use the normal distribution (and hence the Lugannani-Rice formula), symmetric NIG distribution, and asymmetric NIG distribution as the base distributions. We discover that the SPA based on symmetric NIG is quite robust to errors in the computed root, $v$, of the cubic equation given in (\ref{eqn:cubic-cond-sym-NIG}). When the computed value of $v$ becomes greater than $-1$,  symmetric NIG is inapplicable as given by \textbf{Theorem~3}. However, we observe that arbitrarily assuming $v = -1 -\epsilon$, where $\epsilon$ is some small positive number, we can still get reasonable results that  are comparable to SPA based on normal distribution. Hence, in our numerical results, whenever the required condition of SPA based on symmetric NIG failed, $v$ was taken to be $v=-1.000001$. 

\subsection{Uncertainty due to Fading and Number of Interferers}

\begin{figure}[t]
\begin{minipage}{0.5\textwidth}
\begin{center}
	\includegraphics[width=\textwidth]{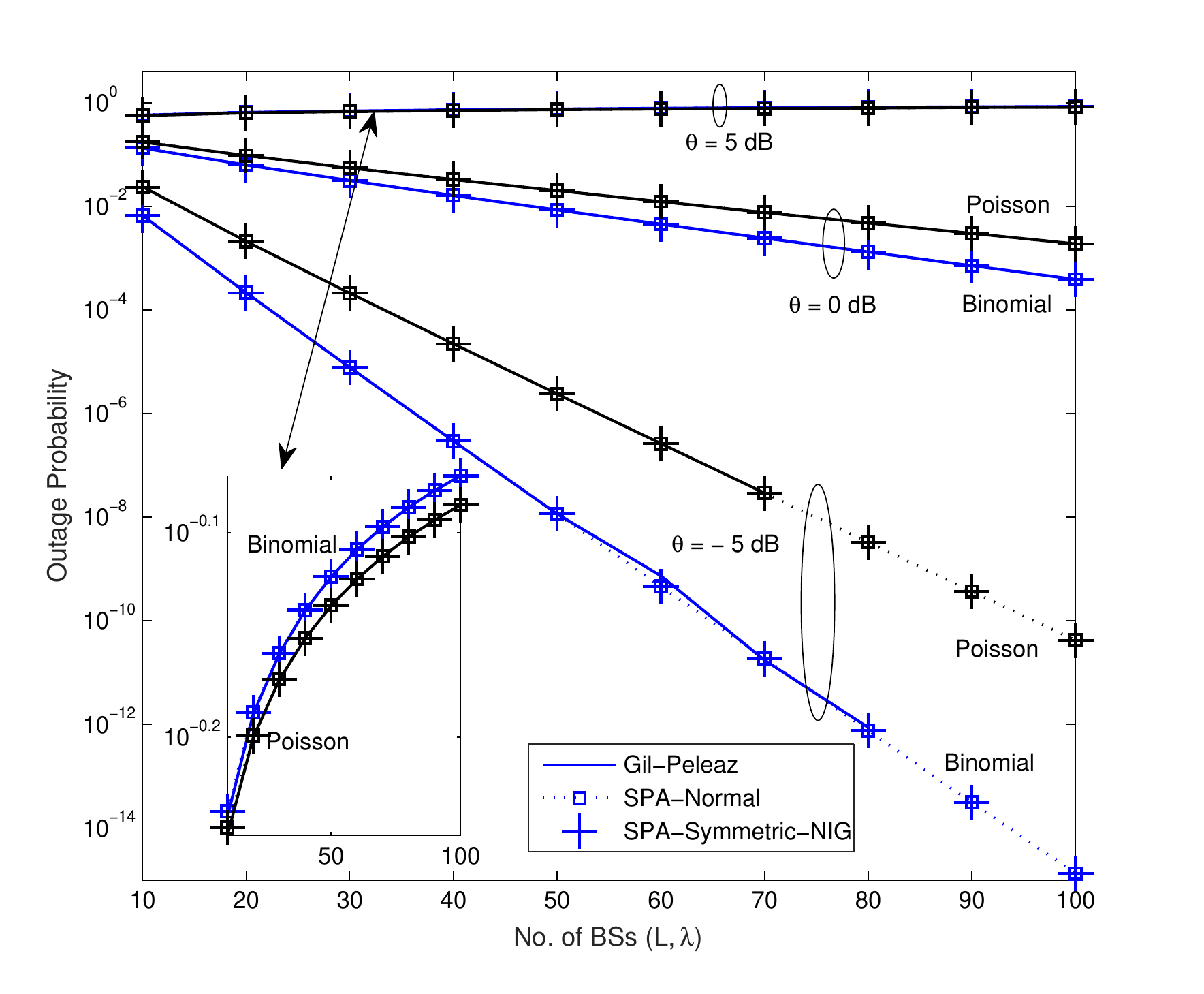}
	\caption{Case A: SIR outage vs. number of BSs.}
	\label{fig:caseA-intensity}
 \end{center}
 \end{minipage}
\hfill
\begin{minipage}{0.5\textwidth}
\begin{center}
	\includegraphics[width=\textwidth]{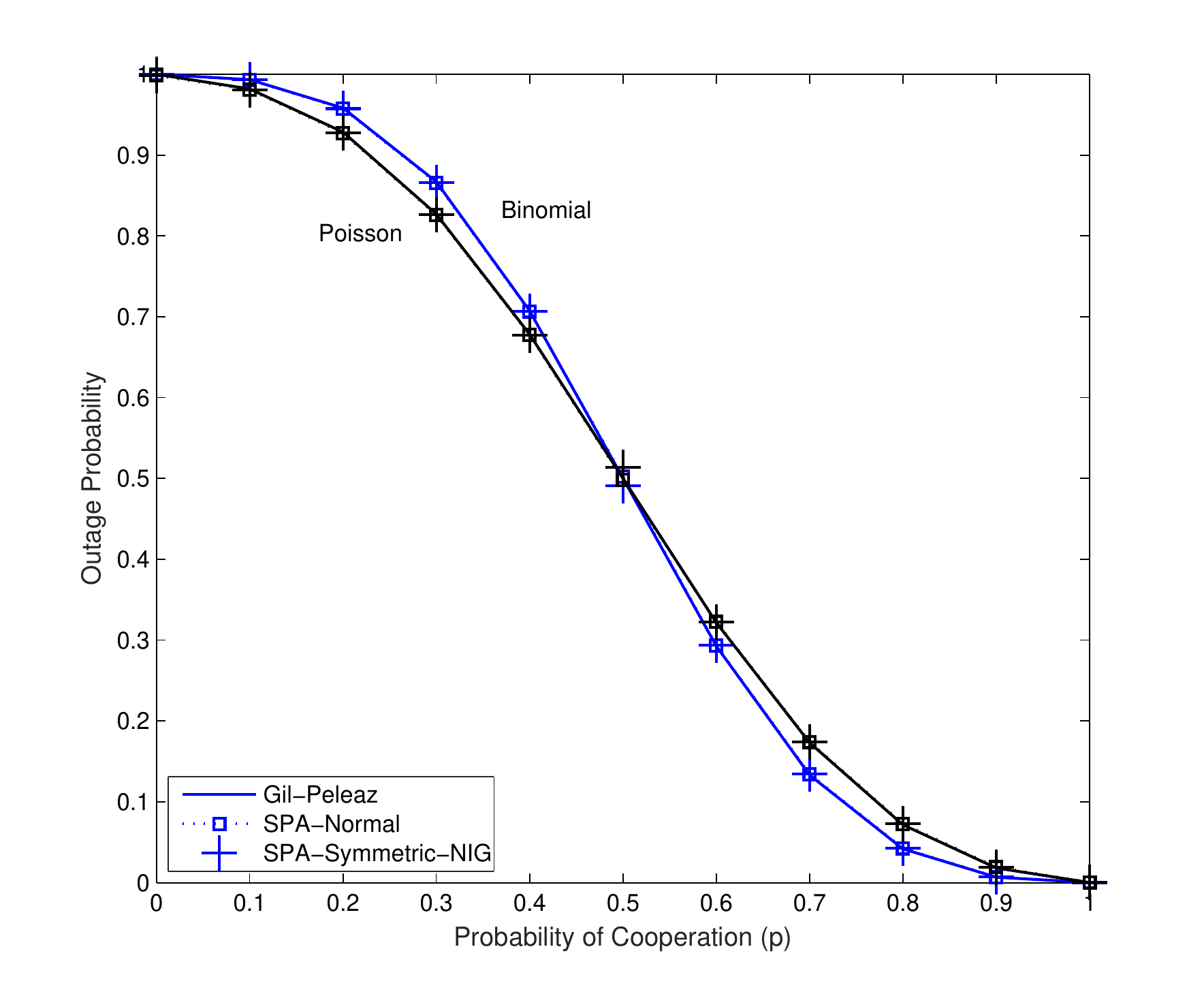}
	\caption{Case A: SIR outage vs. cooperation probability.}
	\label{fig:caseA-coop-prob}
 \end{center}
  \end{minipage}
\end{figure}

In Fig. \ref{fig:caseA-intensity} and Fig. \ref{fig:caseA-coop-prob} we examine the case where uncertainty is due to Poisson and binomial aggregation of interferers coupled with Nakagami-$m$ fading channels. We assume the Nakagami-$m$ fading parameters to be $\alpha = 1$ and $\beta = 1$. For Poisson case, $\lambda_1 = p \lambda$ and $\lambda_2 = q \lambda$, where $p+q = 1$. The saddle point is computed from (\ref{eqn:saddle-pt-poisson-nakagami}) for Poisson aggregation and (\ref{eqn:saddle-pt-binomial-rayleigh}) for binomial aggregation. For both these cases, we only use SPA with symmetric NIG, since both the skewness and kurtosis tend to be quite small (less than five). 

In Fig. \ref{fig:caseA-intensity}, we plot the SIR outage versus the total number of BSs ($\lambda$ in case of Poisson and $L$ in case of binomial).  In both cases, we assume the probability of cooperation to be $p=0.7$. We see that as the number of BSs increases, the SIR outage decreases, when the target SIR threshold $\theta$ is $0$ dB and $-5$ dB. However, when threshold is $\theta = 5$ dB, the SIR outage tends to increase as the number of BSs increases. To understand this reversal in trend, note that the mean $\mu(\Omega) = (-1) \Kcal'_\Omega(0)$. For Poisson case, $\mu(\Omega) = \frac{\alpha \lambda}{\beta}(p - \theta q) \gtrless 0$ is equivalent to $\theta \gtrless \frac{p}{q}$. The mean of $\Omega$ increases or decreases with respect to $\lambda$ depending on the sign of $p - \theta q$. For $p=0.7$, this gives $\theta \gtrless 3.67$ dB. Thus, for $\theta = 5$ dB, increasing $\lambda$ increases $\mu(\Omega)$, thus increasing the CCDF $Q_\Omega(0)$ and hence the outage. Similar reason can be given for the binomial case.

Fig. \ref{fig:caseA-coop-prob} examines the SIR outage as a function of the probability of cooperation $p$. In this plot, we assume $\theta = 0$ dB and $\lambda = L = 10$. We see that as $p$ increases, the SIR outage decreases.

In both Figs. \ref{fig:caseA-intensity} and \ref{fig:caseA-coop-prob}, the accuracy of SPA-based results is comparable to that of the results obtained by direct numerical integration of Gil-Pelaez formula. In Fig. \ref{fig:caseA-intensity}, for $\theta = -5$ dB, we also notice that when the outage probability becomes very small (less than $10^{-8}$), the direct numerical integration of Gil-Peleaz's formula starts to lose its numerical precision and becomes unstable, while the SPA gives consistent results. This happens when the value of the integral is smaller than the tolerable error threshold set for the numerical integration scheme. This is certainly an area where SPA excels direct numerical evaluations. We also note that the results given by SPA based on normal distribution and SPA based on symmetric NIG are consistent. This is not surprising, since the normal distribution is a limiting case of the symmetric NIG distribution.

\subsection{Uncertainty due to Aggregate Interference and Distance}

\begin{figure}[t]
\begin{minipage}{0.5\textwidth}
\begin{center}
	\includegraphics[width=\textwidth]{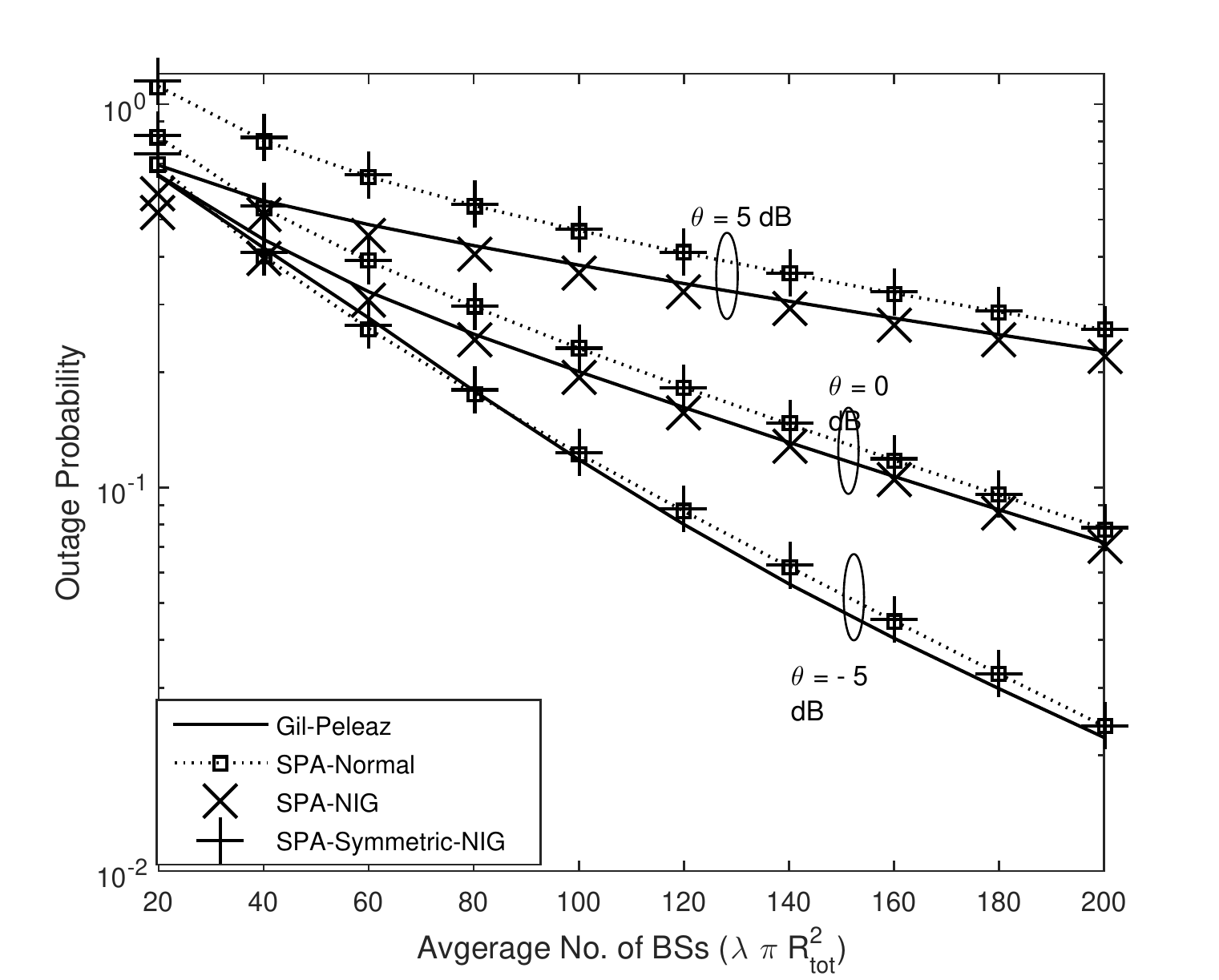}
	\caption{Case B: SIR outage vs. average number of BSs. }
	\label{fig:caseB-intensity}
 \end{center}
 \end{minipage}
\hfill
\begin{minipage}{0.5\textwidth}
\begin{center}
	\includegraphics[width=\textwidth]{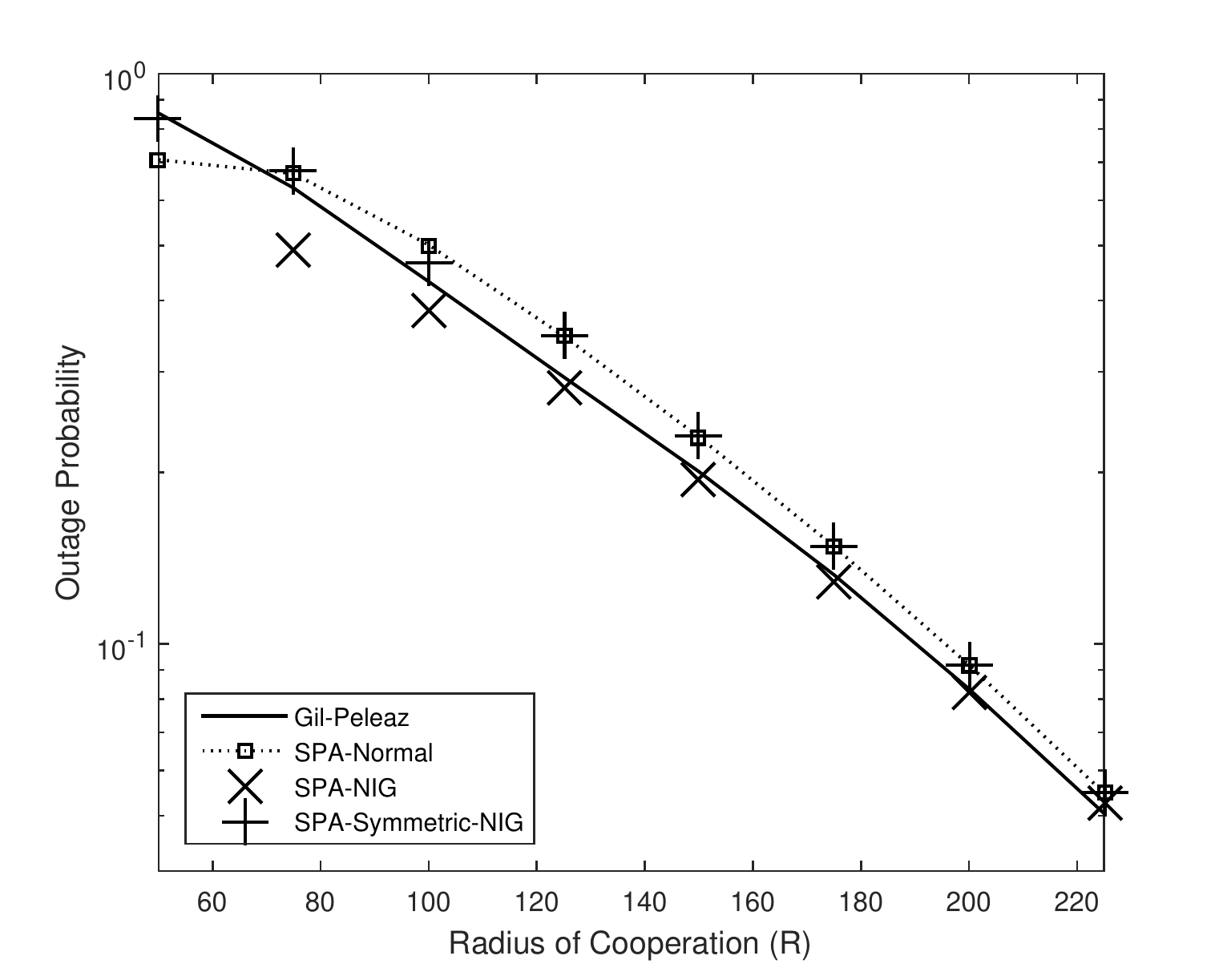}
	\caption{Case B: SIR outage vs. radius of cooperation at $\theta = 0$ dB and $N=100$. }
	\label{fig:caseB-coop-radius}
 \end{center}
  \end{minipage}
\end{figure}

In Figs. \ref{fig:caseB-intensity} and \ref{fig:caseB-coop-radius}, we examine the case where there is uncertainty due to aggregate interference and distance. For these plots, the radius of exclusion region is $a = 30$~m while the radius of cooperation is $R = 150$~m. The path-loss exponent is $\alpha = 4$. The BSs are scattered in a uniform random manner over a total area of radius $R_{tot}=1000$~m such that the average number of BSs in the total area is fixed. The transmit power of the BSs is taken to be $P = 0$ dB. 

In Fig. \ref{fig:caseB-intensity}, we plot the SIR outage versus the average number of BSs in the total area $\pi R_{tot}^2$. Overall, we observe that the outage decreases as the average number of BSs increase. However,  when $\theta$ is $0$ dB and $5$ dB, the SPA based on NIG distribution is clearly more accurate compared the SPA based on normal and symmetric-NIG distributions. Further, both SPA based on normal and symmetric-NIG distributions give similar results. The SPA-based technique is less accurate for smaller number of BSs. For large number of BSs, $\Omega$ becomes more Gaussian, thus both normal and symmetric-NIG based SPA become more accurate at higher number of BSs. For $\theta = -5$ dB, the required condition for SPA based on NIG is not satisfied. Nevertheless, the SPA based on normal and symmetric NIG is observed to yield reasonably accurate results compared to the direct numerical integration of the Gil-Peleaz formula.

In Fig. \ref{fig:caseB-coop-radius}, we plot the SIR outage probability versus the radius of cooperation $R$. For this plot, we assume $\theta = 0$ dB, the average number of BSs as $100$, and the radius of exclusion region $a = 30$~m. Overall, we observe that as the radius of cooperation increases, the SIR outage decreases. We again observe that SPA based on NIG distribution is more accurate than the SPA based on normal and symmetric NIG distributions. The SPA based on all three base distribution become less accurate  when the $R$ is closer to $a$. However, for larger values of $R$, the SPA yields more accurate results.

\section{Conclusion}

We  have proposed a SPA method based on Wood-Booth-Butler formula to calculate the SINR outage probability of any wireless system whose SINR can be modeled in the form $\frac{\sum_{i=1}^M X_i}{\sum_{i=1}^N Y_i +1}$.  
The general approach allows us to deal with distributions with heavy skewness and tails in a more flexible manner. We have exploited asymmetric and symmetric  NIG distribution as well as normal distribution as base distributions for the SPA framework. Numerical results have been presented to check the accuracy of the proposed outage approximation methods. The  NIG based SPA yields more precise and accurate results whenever the sufficient conditions are met. If this is not the case, then we recommend SPA based on normal or symmetric NIG distribution at the expense of reduced accuracy. The framework can be extended to consider the COMP transmission for different diversity combining techniques such as selection combining or equal gain combining. Moreover, the case of $k$-nearest cooperating BSs can also be investigated.


\appendices
 
\section*{Appendix A}
\renewcommand{\theequation}{A.\arabic{equation}}
\setcounter{equation}{0}

\subsection{Finding $\breve{s}(z)$}
We have the CGF of NIG distribution and its first derivative as:
\begin{align}
 \Lcal_Z(s) &= \mu s + \delta (\gamma - \sqrt{\alpha^2 - (\beta+s)^2}), \\
 \Lcal'_Z(s) &= \mu + \frac{ \delta(s + \beta)}{\sqrt{\alpha^2 - (s+\beta)^2}}.
 \end{align}
 
 The saddle point equation $\Lcal'_Z(\breve{s}) = z$ becomes
 \beq 
 \mu + \frac{\delta(\breve{s}+\beta)}{\sqrt{\alpha^2 - (\breve{s}+\beta)^2}} = z.
 \label{eqn:saddle-pt-NIG}
 \eeq
  
 Solving for $\breve{s}(z)$, we have 
 \begin{align}
\quad &\delta^2 (\breve{s}+\beta)^2 = (z - \mu)^2 (\alpha^2 - (\breve{s} + \beta)^2)^2 \nonumber \\ 
\mathrm{or,} \quad & [\delta^2 + (z - \mu)^2](\breve{s} + \beta)^2 = \alpha^2(z - \mu)^2 \nonumber \\ 
\therefore \quad & \breve{s} = - \beta + \frac{\alpha(z-\mu)}{\sqrt{\delta^2 + (z - \mu)^2}}. \label{eqn:breve-s-NIG}
 \end{align}
 
 \subsection{Finding $\hat{z}(x)$}
Letting $c = \Kcal_X(\hat{t}) - \hat{t}x$, we solve for $\hat{z}(x)$ in (\ref{eqn:coincidence-CGF}) as
 \begin{align*}
 & \Lcal_Z(\breve{s}) - \hat{z} \breve{s} = c \\
\mathrm{or,} \quad  & \mu \breve{s} + \delta (\gamma - \sqrt{\alpha^2 - (\beta + \breve{s})^2}) - \hat{z} \breve{s} = c. 
 \end{align*}
Substituting the expression for $\breve{s}$ found in  (\ref{eqn:breve-s-NIG}), and letting $y = \hat{z} - \mu$, we have
\begin{align*}
& \delta \left[ \gamma - \sqrt{\alpha^2 - \frac{\alpha^2 y^2}{\delta^2 + y^2}} \right] - \quad y\left[ - \beta + \frac{\alpha y}{\sqrt{\delta^2 + y^2}}\right]  = c \\
\mathrm{or,} \quad & (\delta \gamma - c) + \beta y - \alpha \sqrt{\delta^2 + \alpha y^2} = 0
\end{align*}
Let $d = \delta \gamma - c$, then the above equation can be written as 
\beq 
(d + \beta y)^2 = \alpha^2 (\delta^2 + y^2).
\label{eqn:y-implicit}
\eeq
We see that (\ref{eqn:y-implicit}) is a quadratic equation. Expanding the square term and re-arranging, we get the equation in standard quadratic form, which we can solve for $y$: 
\begin{align*}
&(\alpha^2 - \beta^2) y^2 - 2 d \beta y + (\alpha^2 \delta^2 - d^2) = 0 \\
\therefore \quad & y = \frac{d\beta \pm \alpha\sqrt{d^2 + \beta^2 \delta^2 - \alpha^2\delta^2}}{\alpha^2 - \beta^2}
\end{align*}
Recalling that $\gamma = \sqrt{\alpha^2 - \beta^2}$, we have $y = \frac{d\beta \pm \alpha\sqrt{d^2 -  \gamma^2 \delta^2}}{\gamma^2}$. Now, putting back the value of $d = \delta \gamma - c$, we obtain $y =  \frac{(\delta \gamma - c) \beta \pm \alpha \sqrt{c^2 - 2\gamma \delta c}}{\gamma^2}$. Putting back the value of $y = \hat{z} - \mu$, we have 
\beq
\hat{z} = \mu +   \frac{(\delta \gamma - c) \beta \pm \alpha \sqrt{c^2 - 2\gamma \delta c}}{\gamma^2}.
\label{eqn:z-NIG}
\eeq

\subsection{Finding Parameters of NIG Distribution}
In order to select the parameters of the NIG distribution, since (\ref{eqn:WBB-formula}) is independent of location and scaling, we first set the mean $\mu$ and scaling parameter $\delta$ to be $\mu = 0$, and $\delta = 1$.

We now need to solve for $\alpha$ and $\beta$, which we can do so using moment matching. Since the saddle point equations (\ref{eqn:saddle-pt-NIG}) gives $\frac{(\breve{s}+\beta)}{\sqrt{\alpha^2 - (\breve{s}+\beta)^2}} = \frac{z - \mu}{\delta}$, we can express the first few derivatives of $\Lcal_Z(s)$ as 
\begin{align}
\Lcal_Z''(\breve{s}) &= \frac{\delta}{\breve{s} + \beta}\left[ \left(\frac{z-\mu}{\delta}\right)^3 + \left(\frac{z-\mu}{\delta}\right) \right],  \label{eqn:L-ii} \\
\Lcal_Z'''(\breve{s}) &= \frac{3\delta}{(\breve{s} + \beta)^2}\left[ \left(\frac{z-\mu}{\delta}\right)^5 + \left(\frac{z-\mu}{\delta}\right)^3 \right], \\
\Lcal_Z^{iv}(\breve{s}) &= \frac{3\delta}{(\breve{s} + \beta)^3}\left[ 5 \left(\frac{z-\mu}{\delta}\right)^7 + 6 \left(\frac{z-\mu}{\delta}\right)^5 + \left(\frac{z-\mu}{\delta}\right)^3 \right]. \label{eqn:L-iv}
\end{align}

Since we have selected $\mu=0$ and $\delta=1$, these derivatives further simplify to
\begin{align*}
\Lcal_Z''(\breve{s}) &= \frac{1}{\breve{s} + \beta}( z^3 + z ),  \\
\Lcal_Z'''(\breve{s}) &= \frac{3}{(\breve{s} + \beta)^2}( z^5 + z^3 ), \\
\Lcal_Z^{iv}(\breve{s}) &= \frac{3}{(\breve{s} + \beta)^3}( 5 z^7 + 6 z^5 + z^3).
\end{align*}

\subsubsection{Finding $\alpha$}
Let the skewness of $X$ at $\hat{t}$ be $\frac{\Kcal_X'''(\hat{t})^2}{\Kcal_X''(\hat{t})^3} = \eta$. Matching the skewness, we have
\begin{align*}
\eta = \frac{\Lcal_Z'''(\breve{s})^2}{\Lcal_Z''(\breve{s})^3} =  \frac{1}{\breve{s} + \beta} \frac{9z^3}{z^2+1}.
 \end{align*} 
 From (\ref{eqn:breve-s-NIG}), after putting the values of $\mu = 0$ and $\delta = 1$, we have $\breve{s} + \beta = \frac{\alpha z}{\sqrt{1 + z^2}}$. Putting this expression for $\breve{s} + \beta$ in the above equation, we have
 \begin{align}
 \eta = \frac{\sqrt{1 + z^2}}{\alpha z} \frac{9z^3}{z^2+1} \qquad  \Rightarrow \qquad  \qquad \alpha = \frac{9 z^2}{\eta \sqrt{1 + z^2}}. \label{eqn:alpha-skew}
\end{align}

Similarly, let the kurtosis of $X$ at $\hat{t}$ be $\frac{\Kcal_X^{iv}(\hat{t})}{\Kcal_X''(\hat{t})^2} = \rho$. Matching the kurtosis, we have
\begin{align*}
\rho  = \frac{\Lcal_Z^{iv}(\breve{s})}{\Lcal_Z''(\breve{s})^2} = \frac{3z(5z^2+1)}{(\breve{s}+\beta)(z^2 + 1)}.
\end{align*}

As before, putting $\breve{s} + \beta = \frac{\alpha z}{\sqrt{1 + z^2}}$, we have
\begin{align}
\rho = \frac{3z(5z^2 + 1)}{z^2+1} \frac{\sqrt{1+z^2}}{\alpha z} \qquad \Rightarrow \qquad \alpha = \frac{3 (5z^2 +1)}{\rho \sqrt{1+z^2}}. \label{eqn:alpha-kurt}
\end{align}

Equating (\ref{eqn:alpha-skew}) and (\ref{eqn:alpha-kurt}), we have the expression for $z$ in terms of $\eta$ and $\rho$ as 
\begin{align*}
 & \frac{9 z^2}{\eta \sqrt{1 + z^2}} = \frac{3 (5z^2 +1)}{\rho \sqrt{1+z^2}} \\ \mathrm{or,} & \quad  5 + \frac{1}{z^2} = \frac{3\rho}{\eta} \\ 
\mathrm{or,} & \quad  z = \left( \frac{3\rho}{\eta} - 5 \right)^{-1/2}. \nonumber
\end{align*}

Taking the value of $z$ to be the same sign as $\hat{t}$, we have a much simpler, alternative expression for $\hat{z}$ purely in terms of $\eta$ and $\rho$:
\beq
\hat{z} = \mathrm{sgn}(\hat{t}) \left( \frac{3\rho}{\eta} - 5 \right)^{-1/2}. \label{eqn:z-skew-kurt}
\eeq

From (\ref{eqn:z-skew-kurt}) and (\ref{eqn:alpha-skew}), we obtain
\begin{align}
 \alpha &= \frac{9}{\eta} \frac{1}{\frac{3\rho}{\eta} - 5} \frac{1}{\sqrt{1+(\frac{3\rho}{\eta}-5)^{-1}}} =  \frac{9}{\sqrt{(3\rho - 5\eta)(3\rho - 4 \eta)}}. \label{eqn:alpha-skew-kurt}
\end{align}

Thus, we see that skewness and kurtosis uniquely defines $z$ and $\alpha$. 

\subsubsection{Finding $\beta$}
Substituting $\mu = 0$ and $\delta = 1$ in (\ref{eqn:y-implicit}), we have $(d+\beta z)^2 = \alpha^2(1 + z^2)$. Also, we have $d = \delta \gamma - c = \sqrt{\alpha^2 - \beta^2} - c$, where $c = \Kcal_X(\hat{t}) - \hat{t}x$. Taking square root on both sides, we have
\begin{align*}
& d + \beta z = \alpha \sqrt{1+z^2} \\
\mathrm{or,}\quad & \sqrt{\alpha^2 - \beta^2} - c = -\beta z + \alpha \sqrt{1+z^2} \\
\mathrm{or,}\quad & \sqrt{\alpha^2 - \beta^2} = c -\beta z + \alpha \sqrt{1+z^2} \\
\therefore \quad & \alpha^2 - \beta^2 = (c -\beta z + \alpha \sqrt{1+z^2})^2.
\end{align*}

Let $e = c + \alpha \sqrt{1+z^2}$, such that we have
\begin{align*}
& \alpha^2 - \beta^2 = (e -\beta z)^2 \\
\mathrm{or,}\quad  & (1+z^2) \beta^2  - 2 e z \beta + (e^2 - \alpha^2) = 0 \\
\therefore \quad & \beta = \frac{e z \pm \sqrt{\alpha^2 + \alpha^2 z^2 -e^2 }}{1+z^2}.
\end{align*}

Taking the same sign as the skew of $X$, we have
\beq
\beta = \frac{e z + \mathrm{sgn}(\Kcal'''_X(\hat{t})) \sqrt{\alpha^2(1 +z^2) -e^2}}{1+z^2}.
\eeq

\subsection{Sufficient Conditions}
For $\beta$ to be real, it is sufficient that $\alpha^2(1 +z^2) -e^2 \geq 0$. Substituting the expression for $e$, we have
\begin{align}
\alpha^2(1 +z^2) -e^2 =\; &  \alpha^2(1+z^2) - (c+\alpha\sqrt{1+z^2})^2 \nonumber \\ 
=\; &  - c (c + \alpha \sqrt{1+z^2} )  \nonumber \\
=\; & - c \left(c + \frac{9}{\sqrt{(3\rho - 5\eta)(3\rho - 4\eta)}} \sqrt{1+\frac{\eta}{3\rho - 5\eta}}\right)  \nonumber \\
=\; & - c \left(c + \frac{9}{3\rho - 5\eta}\right). \label{eqn:beta-positive-condition}
\end{align}

Since both $\rho>0$ and $\eta>0$, the condition $\rho \geq \frac{5}{3}\eta$ implies that both $3\rho - 5 \eta$ and $3\rho - 4\eta$ are non-negative. This ensures that the values of $\hat{z}$ and $\alpha$ are real in (\ref{eqn:z-skew-kurt}) and (\ref{eqn:alpha-skew-kurt}) respectively. 

Assuming that $3\rho - 5\eta \geq 0$, if $c>0$ in (\ref{eqn:beta-positive-condition}), the non-negativity condition for $\beta$ is violated. However, if $c<0$, then $-|c| + \frac{9}{3\rho - 5\eta} \geq 0$ for non-negativity. This reduces to $\rho \leq  \frac{5}{3} \eta + \frac{3}{|c|}$. Thus, combining the non-negativity conditions for $z$, $\alpha$, and $\beta$, we have the sufficient conditions
\[ c < 0 \quad  \mathrm{and} \quad 0 \leq \rho - \frac{5}{3} \eta \leq \frac{3}{|c|}. \]


\section*{Appendix B}
\renewcommand{\theequation}{B.\arabic{equation}}
\setcounter{equation}{0}
\renewcommand{\thesubsection}{\Alph{subsection}}
\setcounter{subsection}{0}

\subsection{Finding Parameters of Symmetric NIG Distribution}
For symmetric NIG distribution, $\beta = 0$. We will further assume that $\mu = 0$. The task is to find suitable $\alpha$ and $\delta$. Using the expressions for $\Lcal^{iv}_Z(\breve{s})$ and $\Lcal'''_Z(\breve{s})$ from (\ref{eqn:L-ii}) and (\ref{eqn:L-iv}), respectively, and putting $w = \frac{z-\mu}{\delta} = \frac{z}{\delta}$, we have $\rho = \frac{3w(5w^2+1)}{\delta(s+\beta)(w^2+1)}$. Substituting $s+\beta = \frac{\alpha (z - \mu)}{\sqrt{\delta^2 + (z -\mu)^2}} = \frac{\alpha w}{\sqrt{1+w^2}}$, in the expression for $\rho$, we have
\begin{align}
\rho &= \frac{3(5w^2+1)}{\alpha \delta \sqrt{1+w^2}}. 
\label{eqn:rho-sym-NIG}
\end{align}

We have from (\ref{eqn:y-implicit}), after substituting $\mu=0$, $(d+\beta z)^2 = \alpha^2 (1+z)^2)$. Here $d = \delta \gamma - c$. Since $\beta=0$, we have $\gamma = \sqrt{\alpha^2 - \beta^2} = \alpha$. Thus we have, after taking square root on both sides,
\begin{align}
& \delta \gamma - c = \alpha \sqrt{\delta^2 + z^2} \nonumber \\
& \delta \alpha - \alpha \delta \sqrt{1+w^2} = c  \nonumber \\
\therefore \quad &  \alpha \delta = \frac{c}{1 - \sqrt{1+w^2}}. \label{eqn:alpha-delta-sym-NIG}
 \end{align} 
 
Comparing (\ref{eqn:alpha-delta-sym-NIG}) and (\ref{eqn:rho-sym-NIG}), we obtain $ \frac{3(5w^2+1)}{\rho\sqrt{1+w^2}} = \frac{c}{1-\sqrt{1+w^2}}$. Let $v = \sqrt{1+w^2}$, then the above expression can be simplified to the following cubic equation
 \begin{align}
 & \frac{3(5v^2 - 4)}{\rho v} = \frac{c}{1-v} \nonumber \\
\therefore \quad & 5 v^3 - 5v^2 + \left(\frac{\rho c}{3}-4\right) v + 4 = 0. \label{eqn:cubic-sym-NIG}
 \end{align}

  We can solve (\ref{eqn:cubic-sym-NIG}) for $v$ after which we have from (\ref{eqn:alpha-delta-sym-NIG})
\beq 
\alpha \delta = \frac{c}{1+v}. 
\label{eqn:alpha-delta-v-sym-NIG}
\eeq 

Clearly, using (\ref{eqn:alpha-delta-v-sym-NIG}) we can select the values of $\alpha$ and $\delta$ in many ways. One simple possibility is to take $\alpha = \delta$ because for this parametrization, as $\alpha \rightarrow \infty$, NIG tends to standard normal distribution. Thus, we have  
\beq
\alpha = \delta = \sqrt{\frac{c}{1+v}}.
\eeq
Finally, since $w = (z-\mu)/\delta$, where $\mu =0$, we have 
\begin{align}
z = \delta w = \sqrt{\frac{c}{1+v}} \cdot \sqrt{v^2 - 1} = \sqrt{c(v-1)}. \label{eqn:z-sym-NIG}
\end{align}

\subsection{Sufficient Condition}
Here root $v$ of the cubic polynomial (\ref{eqn:cubic-sym-NIG}) should be selected such that $c(v-1) > 0$ in (\ref{eqn:z-sym-NIG}) so as to make $z$ real. Similarly, 
$\frac{c}{1+v} > 0$ in (\ref{eqn:alpha-delta-v-sym-NIG}) so that $\alpha = \delta >0$. 

When $c < 0$, the above two positivity conditions imply that $v - 1 <0$ and $1+v <0$. Thus, $v$ is to be chosen such that $v < -1$. Similarly, when $c>0$, the above conditions become $v - 1 >0$ and $1+v>0$. This implies that $v > 1$.
\beq 
v \left\{ \begin{array}{lcr}
		< -1, & \mathrm{for} & c < 0, \\ 
		> 1,  & \mathrm{for} & c> 0.
		\end{array} \right.
\label{eqn:sufficient-condition} 
\eeq

Regardless of the value of $\frac{\rho c}{3}-4$, using Descartes' rule of signs, we can conclude that the cubic equation (\ref{eqn:cubic-sym-NIG}) will have exactly one negative real root. Thus, there is no ambiguity in the selection of $v$ when $c<0$. However, it cannot be guaranteed that criteria (\ref{eqn:sufficient-condition}) is always satisfied.


\section*{Appendix C}
\renewcommand{\theequation}{C.\arabic{equation}}
\setcounter{equation}{0}

Following  \cite{Lowen1990}, the integral of the CGF is evaluated as in (\ref{app-d1})
where $\Gamma(a,z) = \int_z^\infty x^{a-1} e^{-x} \ud x$ is the upper incomplete Gamma function. Note that,
using the recurrence relation $\Gamma(a+1,z) = a\Gamma(a,z) + z^a e^{-z}$, the integral is often expressed as \cite[Eqn 19]{Lowen1990} given in (\ref{app-d2}).

\begin{figure*}[t]
\begin{equation}
\int_a^b (e^{-t P r^{-\alpha}} - 1) r \ud r  = \frac{(tP)^{2/\alpha}}{\alpha}\left[ \Gamma\left(-\frac{2}{\alpha},tPb^{-\alpha}\right) - \Gamma\left(-\frac{2}{\alpha},tPa^{-\alpha}\right) \right] - \frac{b^2 - a^2}{2}.
\label{app-d1}
\end{equation}
\hrule
\end{figure*}

\begin{figure*}[t]
\begin{align}
\int_a^b (e^{-t P r^{-\alpha}} - 1) r \ud r & = -\frac{1}{2}\left[ b^2(1-e^{-tPb^{-\alpha}}) - a^2(1-e^{-tPa^{-\alpha}}) \right. \nonumber \\
& \quad + (tP)^{2/\alpha}\left. \left[ \Gamma\left(1-\frac{2}{\alpha},tPb^{-\alpha}\right) - \Gamma\left(1-\frac{2}{\alpha},tPa^{-\alpha}\right) \right] \right].
\label{app-d2}
\end{align}
\hrule
\end{figure*}

In order to differentiate this integral $n$-times with respect to $t$, consider the $n$-th derivative of the first term:
\begin{align*} 
\frac{\ud^n}{\ud t^n} \frac{(tP)^{2/\alpha}}{\alpha} \Gamma\left(-\frac{2}{\alpha},tPb^{-\alpha}\right) = \frac{b^2}{\alpha}  \frac{\ud^n}{\ud t^n} (tPb^{-\alpha})^{2/\alpha} \Gamma\left(-\frac{2}{\alpha},tPb^{-\alpha}\right). 
\end{align*}
Put $g(t) = tPb^{-\alpha} = z$ and $f(z) = z^{2/\alpha} \Gamma(-2/\alpha,z)$. We have the $n$-th derivative of $f(z)$ with respect to $z$ as \cite[Eqn 8.8.16]{Olver2010} $f^{(n)}(z) = (-1)^n z^{2/\alpha - n} \Gamma\left(n-\frac{2}{\alpha}, z\right)$. Similarly, $g'(t) = Pb^{-\alpha}$, while $g^{(n)} = 0$ for all $n \geq 2$. Now, using Faa di Bruno's formula, $\frac{\ud^n}{\ud t^n} f(g(t)) = \sum_{k=1}^n f^{(k)}(z) B_{n,k}(g'(t),0,\ldots,0 ),$
where $B_{n,k}$ is the partial exponential Bell polynomial. Here $B_{n,k}(g'(t),0,\ldots,0 )$ is 0 if $k<n$ and $g'(t)^n$ if $k=n$. Hence, $\frac{\ud^n}{\ud t^n} f(g(t)) = f^{(n)}(z) \cdot (g'(t))^n = (-1)^n b^{-2} P^{2/\alpha} t^{2/\alpha - n} \Gamma\left(n-\frac{2}{\alpha},tPb^{-\alpha}\right)$. Therefore, we have 
\begin{eqnarray}
\frac{\ud^n}{\ud t^n} \frac{(tP)^{2/\alpha}}{\alpha} \Gamma\left(-\frac{2}{\alpha},tPb^{-\alpha}\right)  =  \frac{(-1)^n}{\alpha}  P^{2/\alpha} t^{2/\alpha - n} \Gamma\left(n-\frac{2}{\alpha},tPb^{-\alpha}\right).
\end{eqnarray}

 We will have similar result for $\frac{(tP)^{2/\alpha}}{\alpha} \Gamma\left(-\frac{2}{\alpha},tPa^{-\alpha}\right)$. Lastly, the derivatives of the constant last term will be zero. Putting everything together, we have our desired result.

 
 \section*{Acknowledgments}
The authors would like to thank Prof. Martin Haenggi for helpful discussions. This work was funded by the Natural Sciences and Engineering Research Council of Canada (NSERC).  


\bibliographystyle{IEEE}

 \end{document}